\documentclass[twocolumn]{aastex631}
\usepackage{graphicx}
\usepackage{color}
\usepackage{amsmath}
\usepackage{threeparttable}
\usepackage{multirow}
\usepackage{natbib}
\usepackage{hyperref}
\usepackage{appendix}

\begin{document}

\title{Distinguishing Tidal Disruption Events and Changing-look Active Galactic Nuclei via Variation of Mid-infrared Color}

\author{Yujun Yao}
\affiliation{Department of Physics, Anhui Normal University, Wuhu, Anhui 241002, China}

\author{Jingjing Ye}
\affiliation{Department of Physics, Anhui Normal University, Wuhu, Anhui 241002, China}

\author{Luming Sun}
\affiliation{Department of Physics, Anhui Normal University, Wuhu, Anhui 241002, China}

\author{Ning Jiang}
\affiliation{CAS Key Laboratory for Researches in Galaxies and Cosmology, Department of Astronomy, University of Science and Technology of China, Hefei, Anhui 230026, China}
\affiliation{School of Astronomy and Space Sciences, University of Science and Technology of China, Hefei, 230026, China}

\author{Megan Masterson}
\affiliation{MIT Kavli Institute for Astrophysics and Space Research, Massachusetts Institute of Technology, Cambridge, MA 02139, USA}

\author{Xinwen Shu}
\affiliation{Department of Physics, Anhui Normal University, Wuhu, Anhui 241002, China}

\correspondingauthor{Luming~Sun}
\email{sunluming@ahnu.edu.cn}

\begin{abstract}
  In this work, we propose that the time variation of mid-infrared (MIR) color is a promising probe to distinguish between MIR outbursts induced by tidal disruption events (TDEs) and changing-look active galactic nuclei (CLAGNs).
  With an optically selected sample containing TDEs, ambiguous nuclear transients (ANTs), and CLAGNs, we studied the variation of MIR color ($W1-W2$) after subtracting the quiescent fluxes using NEOWISE-R data.
  The MIR color of TDEs and ANTs turns red faster than CLAGNs during the rising phase, as the color variation rate (CVR) of TDEs and ANTs is generally $\gtrsim0.2$ mag year$^{-1}$, whereas that of CLAGNs is generally $\lesssim0.3$ mag year$^{-1}$.
  This may be caused by the difference between the ultraviolet light curves of TDEs/ANTs and CLAGNs, or be related to no or weak underlying AGN in TDEs/ANTs.
  In addition, TDEs have a redder color than ANTs at the earliest phase.
  Based on CVR, we selected high-probability TDE, ANT, and CLAGN candidates from MIR outbursts in samples of Jiang et al. (2021) and Masterson et al. (2024).
  We found that both samples are mixtures of TDEs/ANTs and CLAGNs.
  For MIR outbursts whose hosts are not Seyfert galaxies, we estimated that $\sim50\%-80\%$ are TDEs and inferred a rate of infrared TDEs of $1.5-2.8\times10^{-5}$ galaxy$^{-1}$ yr$^{-1}$, comparable with that of optical TDEs;
  The rest are CLAGNs, suggesting the presence of weak AGNs that cannot be identified using common diagnoses.
  Our work opens a new door for future classification of infrared selected transients based on only MIR photometric data.
\end{abstract}

% while the latter may be related to the difference in the dust geometry.
%  \textbf{and for MIR outbursts whose hosts are Seyfert galaxies we infrared CLAGNs of $0.8-1.4\times10^{-4}$ galaxy$^{-1}$ yr$^{-1}$, less than optical CLAGNs.}

\section{Introduction} \label{sec:introduction}

In quiescent galaxies, which account for the majority of galaxies, tidal disruption events (TDEs) are powerful probes of supermassive black holes (SMBHs).
A TDE occurs when a star wanders too close to the SMBH for its own gravity to resist the tidal force \citep[e.g.,][]{Rees1988}.
TDEs can be observed through bright flares lasting for months to years \citep[e.g.,][]{vanVelzen2020_review}.
Thanks to the development of time-domain surveys, $\gtrsim100$ TDEs have been discovered.
Most of TDEs are selected in the optical band (optical TDEs) via their bright UV-optical emission \citep[e.g.,][]{vanVelzen2020_review,Hammerstein2023}, while a small number are selected in the X-ray band (X-ray TDEs) via their bright X-ray emission \citep[e.g.,][]{Saxton2020,Sazonov2021}, and some are bright in both bands \citep[e.g.,][]{Guolo2024}.
According to recent sample studies, the incidence rate of optical TDEs is around $3\times10^{-5}$ galaxy$^{-1}$ yr$^{-1}$ from the Zwicky Transient Facility (ZTF) sample \citep[e.g.,][]{Yao2023}, while that of X-ray TDEs is around $1\times10^{-5}$ galaxy$^{-1}$ yr$^{-1}$ from the eROSITA sample \citep[e.g.,][]{Sazonov2021,Grotova2025}.
In addition, the TDE rate is related to the properties of the host galaxy.
For example, post-starburst galaxies have a higher incidence rate of TDEs than normal galaxies \citep[e.g.,][]{French2016,Yao2023}.
Thus, the above-mentioned TDE rates may be the average value of different types of galaxies.

Despite these findings, there is a lack of complete TDE samples because dust-obscured TDEs are hard to detect by optical or X-ray surveys.
Fortunately, the TDE's UV emission heats the surrounding dust to produce infrared (IR) re-radiation, known as IR echo \citep[e.g.,][]{Jiang2016,vanVelzen2016}.
Dust-obscured TDEs can reveal themselves by their IR echoes as IR outbursts, for example, Arp 299-B AT1 \citep{Mattila2018} and AT 2017gbl \citep{Kool2020}.
Thanks to the all-sky mid-infrared (MIR) sky survey of the Wide-field Infrared Survey Explorer \citep[WISE,][]{Wright2010}, hundreds of MIR outbursts in centers of galaxies have been discovered \citep{Jiang2021,Reynolds2022,Masterson2024,Necker2025}.
More than $\sim$80\% of these MIR outbursts were not detected in optical/X-ray bands.
On the other hand, optical TDEs rarely have strong IR emission \citep{Jiang2021_optsample}.
These suggest that optical/X-ray TDEs and MIR outbursts may be two distinct groups with little overlap.
The event rate of MIR outbursts obtained by \citet{Jiang2021} is $\sim5\times10^{-5}$ galaxy$^{-1}$ yr$^{-1}$, while that by \citet{Masterson2024} and \citet{Necker2025} is lower with $\sim2-3\times10^{-5}$ galaxy$^{-1}$ yr$^{-1}$, possibly because the former work did not remove the contribution of AGN outbursts.
These event rates are similar to the rate of optical and X-ray TDEs, implying that optical and X-ray surveys may have missed a significant fraction of TDEs.

However, a key question remains unanswered: are these MIR outbursts TDEs?
Although the authors of previously mentioned works preferred that their samples consist mainly of obscured TDEs, there is doubt \citep[e.g.,][]{Dodd2023} that changing-look active galactic nuclei (CLAGN) contribute to these MIR outbursts significantly.
CLAGNs \citep[see review in][]{Ricci2023} can also produce MIR outbursts \citep[e.g.,][]{Sheng2017,Lyu2022}.
Despite many differences between TDEs and CLAGNs in UV/optical bands in terms of light curves (LCs) and optical spectra \citep[e.g.,][]{Zabludoff2021}, distinguishing their IR echoes remains challenging.
Differences in the UV/optical LCs are blurred by the light travel time effect of the echo, and the features that may distinguishing the two are lost when the emission is reprocessed to the IR band.
To make matters worse, there is a class of outbursts in active galactic nuclei (AGN) that share the characteristics of both TDEs and CLAGNs, for example AT 2016ezh \citep[=PS16dtm,][]{Blanchard2017,Petrushevska2023}, AT 2017bgt \citep{Trakhtenbrot2019}, AT 2018bcb \citep[=ASASSN-18jd,][]{Neustadt2020}, and some others.
These outbursts are referred to as ambiguous nuclear transients (ANTs) as their natures are difficult to determine through UV/optical information: they may be TDEs that occur in AGNs \citep[e.g.,][]{Blanchard2017}, or enhanced accretions onto SMBHs \citep[e.g.,][]{Trakhtenbrot2019}.
Recent MIR spectroscopy of TDEs shows that TDEs generally have strong silicate emission features, different from AGNs, which show absorption or weak emission features \citep{Masterson2025}.
This implies that the silicate feature has the potential to serve as a probe to distinguish TDEs from CLAGNs.
In addition, the nature of MIR outbursts can be distinguished by radio jet \citep[e.g. for Arp 299-B AT1,][]{Mattila2018}, broad emission lines in the near-infrared (NIR) spectra \citep[e.g. for AT 2017gbl,][]{Kool2020} or optical spectra \citep[e.g.,][]{Wang2022_MIRONG}, or shape of X-ray spectrum \citep[e.g.,][]{Masterson2024}.
However, these data are not available for the vast majority of MIR outbursts.
Therefore, there needs to be an effective probe to distinguish between TDEs and CLAGNs based on only MIR photometric data.

\citet{Jiang2017} noticed that the dust temperature in AT 2016ezh was high with a blackbody temperature $T_{\rm BB}\gtrsim2,000$ K initially, and then rapidly dropped to $T_{\rm BB}\sim1,000$ K in 100--300 days.
Although the temperature values here may not be accurate because the dust properties are uncertain, the trend of temperature decline is robust.
Similar rapid drops in dust temperatures have also been observed in other TDEs, ANTs and TDE candidates \citep[e.g.,][]{Mattila2018,Sun2020,Kool2020,Reynolds2022,Hinkle2024,Necker2025}.
However, such drops were not found in CLAGNs\footnote{Although some works claimed a ``redder when brighter'' trend in MIR emission of CLAGNs \citep{Yang2018}, they did not subtract the host galaxy's contribution to IR emission when they calculated the color}.
We guessed that there may be a general difference in the time variation of dust temperature between TDEs and CLAGNs.
The primary purpose of this work is to verify this difference and attempt to distinguish TDEs from CLAGNs using this difference.

This paper is organized as follows.
We collected optically selected outbursts, generated MIR LCs and selected those with strong IR echoes in section 2.
We studied the MIR color variation and proposed a new method for distinguishing TDEs and CLAGNs in section 3.
We applied the method to MIR outburst samples and selected promising TDE and CLAGN candidates in section 4.
We discuss in section 5 and summarize and conclude in section 6.
Thoughout this paper, we used cosmological parameters of $H_0=70$ km s$^{-1}$ Mpc$^{-1}$, $\Omega_m=0.3$ and $\Omega_\Gamma=0.7$.

\section{Optically-selected outbursts and data reduction} \label{sec:opt_sample}

In order to study the variation of MIR colors of TDEs, CLAGNs and ANTs, and to check the disparity between different types, we collected outbursts whose type can be defined by optical observations.
Although in the literature, the term ANT was also used by some authors to refer to different meanings, we will strictly use the term to mean the transient source occurring in an AGN whose nature is being debated.
We sifted the samples to ensure that MIR colors can be measured with small errors.
The selection criteria and the final samples are described in detail in sections \ref{sec:opt_father_sample} to \ref{sec:opt_rising}, and the MIR properties of the sample are displayed in section \ref{sec:opt_mir_property}.

\subsection{MIR-bright TDEs, CLAGNs and ANTs} \label{sec:opt_father_sample}

We collected optical TDEs from the ZTF sample \citep{vanVelzen2021,Hammerstein2023,Yao2023} and other literature.
We applied a redshift cut of $z < 0.35$, the same as \cite{Jiang2021}.
The NEOWISE-R project \citep{Mainzer2014} provides the MIR data of all the objects in the W1 and W2 bands (central wavelengths 3.35 and 4.60 $\mu$m) between its start in December 2013 and the WISE telescope's retirement in July 2024.
We obtained the MIR LCs in the W1 and W2 bands sampled every half a year following \citet{Jiang2021}.
In summary, we queried single-exposure photometries requiring separation of less than 3 arcsecs using NASA/IPAC\footnote{https://irsa.ipac.caltech.edu/frontpage/}.
We removed frames with poor qualities (${\rm qi\_fact}<1$), or affected by charged particle hits (${\rm saa\_sep}<5$), scattered moonlight (${\rm moon\_masked}=1$), or artifacts (${\rm cc\_flags}\neq0$), and binned the data points every half a year according to the WISE survey strategy.
To ensure that the MIR color of the transient can be measured with high precision, we further required that the transient is MIR bright with $m_{\rm min,{\rm W2}}<12.5$, and cause a large amplitude of variation with $\delta{\rm W2}>0.5$ during the NEOWISE-R phase, where $m_{\rm min,{\rm W2}}$ and $\delta{\rm W2}$ are the minimum magnitude in the W2 band and its difference from the maximum value, respectively.
%These criteria are stricter than those of \citet{Jiang2021} ($m_{\rm min,{\rm W2}}>14$, $\delta{\rm W1}>0.5$ or $\delta{\rm W2}>0.5$).
Nine TDEs passed our criteria, including AT 2017gge, AT 2018dyk, AT 2018gn, AT 2019dsg, AT 2019qiz, AT 2020nov, AT 2022upj, AT 2023ugy, and SDSS J1115+0544.
Note that AT 2018dyk was originally considered to be a changing-look LINER because some of its LC features and spectral characteristics are not common in TDEs \citep{Frederick2019}.
However, as its observational data were reanalyzed in more detail and the diversity of TDEs was more fully understood, it was reclassified as TDE \citep{Huang2023,Clark2025}.
In addition, SDSS J1115+0544 was initially reported as a turn-on AGN because the transient remained at a plateau until 600 days after its peak \citep{Yan2019}.
However, follow-up works found that it dimmed within 4--5 years, reclassifying it as a slow-evolved TDE \citep{Wang2022_MIRONG,Zhang2025}.

\citet{Lyu2022} has investigated the MIR properties of 68 CLAGNs reported in the literature.
We started from Lyu's sample but removed AT 2018dyk\footnote{Referred to as SDSS J153308.02+443208.4 in \citet{Lyu2022}} and SDSS J1115+0544.
We generated the MIR LCs of the remaining 66 objects and selected bright objects with large amplitude variability in the same way as we described previously.
Two CLAGNs, NGC 4151 and NGC 7582, are too bright, and their photometries are severely affected by saturated pixels, so we removed them from the sample.
After setting the same criteria as for TDEs, we selected 28 MIR bright CLAGNs with significant MIR variability.
As pointed out by \citet{Sheng2017}, the CLAGNs in the sample have large amplitudes of MIR variability, and hence are likely to originate from changes in the accretion rate, rather than changes in obscuration or other causes.

We collected 26 ANTs\footnote{We removed AT 2017gge and AT 2018dyk in \citet{Hinkle2024} as they were classified as TDEs by us.} from literatures \citep{Frederick2021,Hinkle2024,Wiseman2025}.
We selected the sample using the same criteria and eight ANTs passed, including AT 2016ezh, AT 2017bgt, AT 2018bcb, AT 2019aalc, AT 2019avd, AT 2020agdm, AT 2021loi and AT 2022fpx.

\subsection{Quiescent and outburst states} \label{sec:opt_quiescent_outburst}

We preliminary analysed the MIR LCs to determine the time duration of quiescent and outburst states.

\begin{figure}
\centering
 \includegraphics[scale=0.66]{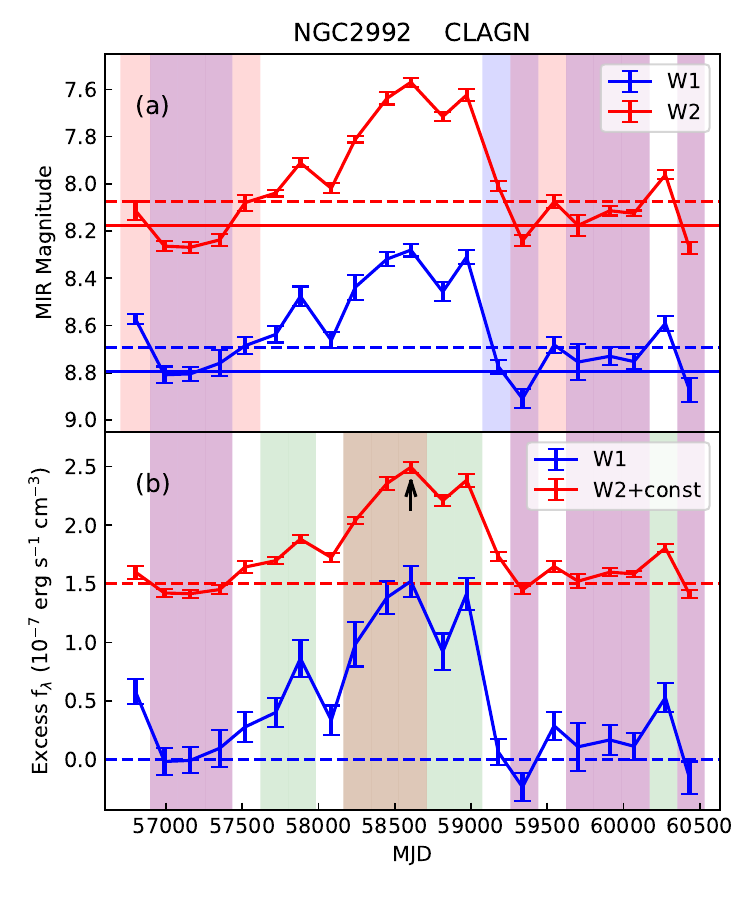}
 \caption{
  \textbf{(a)}: Illumination of selecting quiescent state.
  The horizontal solid lines show the quiescent levels, and the dashed lines show the magnitude thresholds for selecting the quiescent state.
  The shades indicate the final quiescent states: shallow blue for the W1 band, shallow red for the W2 band, and shallow purple for their intersection.
  \textbf{(b)}: Illumination of selecting outburst state (green and orange shades) and the rising phase (orange shade).
  The black arrow labels the peak time.}
  \label{fig:example_quie_outb}
\end{figure}

We selected a quiescent state for each object using an iterative algorithm.
For an LC in a particular band, we took the data point with the largest magnitude and those with a magnitude difference of less than 0.2 from it as the initial quiescent state.
We then calculated the weighted average of the quiescent state's magnitudes (weighted by the reciprocal of the squared of the error) and the standard deviation and added data points that differ by less than 0.1 and less than 1.635 standard deviations from the weighted average to the quiescent state.
We then recalculated the weighted average and standard deviation with the updated quiescent state and attempted to add new data points to the quiescent state until there were no more data points to add.
We calculated the quiescent state for each of the W1 and W2 bands using the above algorithm and finally adopted the intersection of the results of the two bands as the quiescent state of the object.
The results are consistent with visual inspections, as shown in Figure~\ref{fig:example_quie_outb}(a).

We calculated the weighted average of the fluxes in the final quiescent state as the quiescent fluxes.
We then calculated excess fluxes $f_e$ by subtracting the quiescent fluxes from the original fluxes and their errors using the law of propagation of uncertainties.
For data points not in the quiescent state, we finally selected points where the excess fluxes exceed three times their errors in both W1 and W2 bands as an ``outburst state'', as an example shown in Figure~\ref{fig:example_quie_outb}(b).

\begin{figure}
\centering
 \includegraphics[scale=0.66]{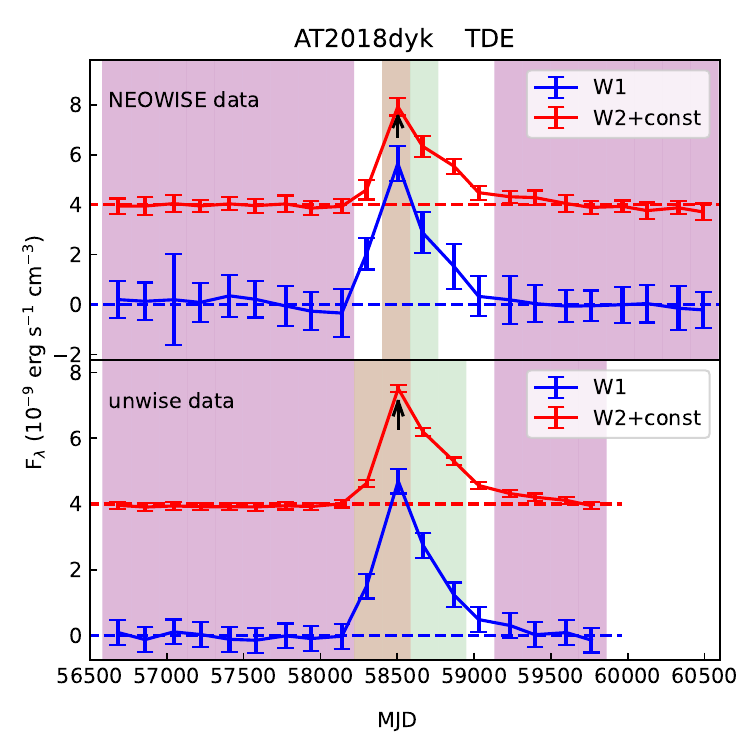}
 \caption{
  The selections for the outburst state and rising phase for AT 2018dyk.
  The upper and lower panels are the results using NEOWISE data and unWISE data, respectively.
  The meaning of labels and colors are the same as Figure~\ref{fig:example_quie_outb}(b).}
\label{fig:example_unwise}
\end{figure}

The above analysis is based on LCs generated by NEOWISE single-exposure photometries.
This method of generating LCs was proved to be reliable \citep[e.g.,][]{Jiang2021}; however, it leads to large magnitude errors for faint MIR sources.
Thus, for objects with a quiescent W2 magnitude $>$11, we generated LCs using time-resolved coadded data provided by the unWISE project \citep{Meisner2018} with an image subtraction method, which was also proved reliable \citep[e.g.,][]{Masterson2024}.
For each object in each band, we made a reference image using the median stack of the quiescent-state images, and then subtracted it from all the images to obtain difference images.
We measured flux and error on each difference image using a PSF model whose center is fixed at the optical position for TDEs and ANTs or at the galaxy center for CLAGNs.
In addition to routine statistical flux error, we considered the systematic error due to image subtraction, which is 2\% of the quiescent PSF flux according to our tests using galaxies with no outbursts.
We adopted the fluxes measured on difference images as excess fluxes $f_e$, and selected the outburst state in the same way as described previously.
In Figure~\ref{fig:example_unwise}, we show an example that the data quality improves by using unWISE data.
The flux errors are smaller, and more data points are selected for the outburst state, so the variation of MIR color can be better studied.
However, because we have not obtained unWISE data in 2023 and beyond, we still use NEOWISE data for transients that started late, for example, AT 2022fpx, AT 2022upj, and AT 2023ugy.

\subsection{Rising phase and further selections} \label{sec:opt_rising}

Based on numerical simulations, \citet{Sun2020} demonstrated that the time variation of MIR color in the rising phase is critical to probe the nature of a MIR outburst.
Therefore, we selected the rising phase of each outburst for the following analysis.

We selected consecutive data points in the outburst state with $t<=t_{\rm peak}$ as the rising phase, where $t_{\rm peak}$ is the peak time in the W2 band, as examples shown in Figure~\ref{fig:example_quie_outb}(b) and Figure~\ref{fig:example_unwise} with orange shades.

We excluded eight CLAGNs because the rising phase may not be completely sampled by NEOWISE-R, which will be described in detail in Appendix \ref{appendix:special}.
The final sample contains 37 objects, including 9 TDEs, 20 CLAGNs, and 8 ANTs.
We list the redshift, the type and the reference in Table~\ref{tab:information_CVR_IC}.

%%% sources with rising phase starts from the first data point (7 CLAGNs):
%%%    3C 390.3, HE1136-2304, Mrk 6, Mrk 926, NGC 2617, NGC 4395, and SDSS J080020.98+263648.8
%%% sources peaks at the last data point (2 TDEs, 3 ANTs and 5 CLAGNs):
%%%    AT2019qiz, AT2023ugy; AT2016ezh, AT2019aalc, AT2021loi; 2MASXJ09483841+4030436, ESO362-G18, IRAS23226-3843, NGC3516, SDSSJ122550.30+510846.3
%%%    ESO362-G18 has only one data point in the rising phase

% AT 2017gge = ATLAS17jrp
% AT 2018gn = ASASSN-18ap
% AT 2016ezh = PS16dtm
% AT 2018bcb = ASASSN-18jd
% AT 2020adgm = ASASSN-20jd
\begin{table*}[!t]
\centering
\caption{Information of the optical transients and parameters of MIR color variation.}
\begin{tabular}{cccccc}
\hline
\hline
Object           & RedShift  & Type  & References & CVR    & IC  \\
\hline
AT2017gge        & 0.066    & TDE   & 1,2         & $-$               & $0.60 \pm 0.03 $ \\
AT2018dyk        & 0.0367   & TDE   & 3,4,5       & $1.11 \pm 0.62 $  & $0.34 \pm 0.32 $ \\
AT2018gn         & 0.0375   & TDE   & 6           & $0.33 \pm 0.04 $  & $0.51 \pm 0.05 $ \\
AT2019dsg        & 0.0512   & TDE   & 7           & $0.66 \pm 0.15 $  & $0.67 \pm 0.05 $ \\
AT2019qiz        & 0.0151   & TDE   & 7           & $0.27 \pm 0.03 $  & $0.60 \pm 0.06 $ \\
AT2020nov        & 0.084    & TDE   & 8           & $0.75 \pm 0.23 $  & $0.31 \pm 0.09 $ \\
AT2022upj        & 0.054    & TDE   & 9           & $0.48 \pm 0.18 $  & $0.52 \pm 0.04 $ \\
AT2023ugy        & 0.106    & TDE   & 10          & $0.46 \pm 0.42 $  & $0.41 \pm 0.18 $ \\
SDSS J1115+0544  & 0.09     & TDE   & 11,12,13    & $0.41 \pm 0.09 $  & $0.68 \pm 0.06 $ \\
AT2016ezh        & 0.0804   & ANT   & 14,15       & $0.37 \pm 0.06 $  & $0.27 \pm 0.14 $ \\
AT2017bgt        & 0.064    & ANT   & 16          & $0.41 \pm 0.18 $  & $0.24 \pm 0.28 $ \\
AT2018bcb        & 0.1192   & ANT   & 17          & $-$               & $0.49 \pm 0.03 $ \\
AT2019aalc       & 0.0356   & ANT   & 18          & $0.35 \pm 0.09 $  & $0.43 \pm 0.07 $ \\
AT2019avd        & 0.0296   & ANT   & 19,20,21,22 & $0.20 \pm 0.04 $  & $0.50 \pm 0.06 $ \\
AT2020adgm       & 0.056    & ANT   & 23          & $0.56 \pm 0.10 $  & $0.24 \pm 0.09 $ \\
AT2021loi        & 0.083    & ANT   & 24          & $0.35 \pm 0.04 $  & $0.30 \pm 0.03 $ \\
AT2022fpx        & 0.073    & ANT   & 25          & $0.32 \pm 0.20 $  & $0.42 \pm 0.30 $ \\
2MASX J0938+0743 & 0.022    & CLAGN & 26          & $0.33 \pm 0.26 $  & $0.40 \pm 0.38 $ \\
2MASS J0948+4030 & 0.0468   & CLAGN & 26          & $0.04 \pm 0.21 $  & $0.59 \pm 0.27 $ \\
IRAS23226-3843   & 0.0359   & CLAGN & 26          & $0.24 \pm 0.16 $  & $0.36 \pm 0.25 $ \\
Mrk1018          & 0.043    & CLAGN & 26          & $0.87 \pm 0.29 $  & $0.36 \pm 0.10 $ \\
NGC0863          & 0.0264   & CLAGN & 26          & $0.08 \pm 0.11 $  & $0.76 \pm 0.16 $ \\
NGC1566          & 0.005    & CLAGN & 26          & $-0.55 \pm 0.29 $ & $1.09 \pm 0.15 $ \\
NGC2992          & 0.0077   & CLAGN & 26          & $0.19 \pm 0.25 $  & $0.66 \pm 0.22 $ \\
NGC3516          & 0.0088   & CLAGN & 26          & $-0.02 \pm 0.12 $ & $0.64 \pm 0.23 $ \\
NGC4388          & 0.0084   & CLAGN & 26          & $-0.01 \pm 0.28 $ & $1.28 \pm 0.25 $ \\
SDSS J0813+4608  & 0.0538   & CLAGN & 26          & $0.09 \pm 0.12 $  & $0.50 \pm 0.12 $ \\
SDSS J0817+1012  & 0.0458   & CLAGN & 3,26        & $-0.02 \pm 0.09 $ & $0.76 \pm 0.22 $ \\
SDSS J0829+4154  & 0.1263   & CLAGN & 26          & $0.11 \pm 0.12 $  & $0.19 \pm 0.24 $ \\
SDSS J0915+4814  & 0.1005   & CLAGN & 3,26        & $0.12 \pm 0.27 $  & $0.55 \pm 0.38 $ \\
SDSS J1133+6701  & 0.0397   & CLAGN & 3,26        & $0.13 \pm 0.06 $  & $0.47 \pm 0.09 $ \\
SDSS J1225+5108  & 0.1679   & CLAGN & 3,26        & $0.13 \pm 0.26 $  & $0.50 \pm 0.11 $ \\
SDSS J1625+2415  & 0.0503   & CLAGN & 26          & $0.06 \pm 0.21 $  & $0.65 \pm 0.16 $ \\
WISEA J0847+1824 & 0.0848   & CLAGN & 26          & $0.16 \pm 0.16 $  & $0.53 \pm 0.33 $ \\
WISEA J1003+3525 & 0.1189   & CLAGN & 26          & $-0.06 \pm 0.06 $ & $0.52 \pm 0.10 $ \\
WISEA J1545+1709 & 0.0483   & CLAGN & 26          & $-0.05 \pm 0.10 $ & $0.71 \pm 0.15 $ \\
WISEA J1545+2511 & 0.117    & CLAGN & 26          & $0.03 \pm 0.13 $  & $0.55 \pm 0.22 $ \\
\hline
\end{tabular}
\begin{tablenotes}
   \item References: 1. \cite{Wang2022_AT2017gge}; 2. \cite{Onori2022}; 3. \cite{Frederick2019}; 4. \cite{Huang2023}; 5. \cite{Clark2025}; 6. \cite{Wang2024_AT2018gn}; 7. \cite{vanVelzen2021}; 8. \cite{Earl2025}; 9. \cite{Newsome2024}; 10. \cite{Yao2023_AT2023ugy}; 11. \cite{Yan2019}; 12. \cite{Wang2022_MIRONG}; 13. \cite{Zhang2025}; 14. \cite{Blanchard2017}; 15. \cite{Petrushevska2023}; 16. \cite{Trakhtenbrot2019}; 17. \cite{Neustadt2020}; 18. \cite{Veres2024_AT2019aalc}; 19. \cite{Malyali2021}; 20. \cite{Frederick2021}; 21. \cite{Chen2022}; 22. \cite{Wang2024_AT2019avd}; 23. \cite{Kosec2023}; 24. \cite{Makrygianni2023}; 25. \cite{Wiseman2025}; 26. \cite{Lyu2022}.
\end{tablenotes}
\label{tab:information_CVR_IC}
\end{table*}

\subsection{MIR properties of the samples} \label{sec:opt_mir_property}

In this subsection, we show the MIR properties of the optically selected transients, including $m_{\rm min,W2}$, $\delta{\rm W2}$, the peak W2 luminosity $L_{\rm max,W2}$ after quiescent level subtracted, and the rise time $t_{\rm rise}$.
We calculated $L_{\rm max,W2}$ using luminosity distance $D_L$ converted from redshift $z$ if $z>0.01$, otherwise we used the median value of $D_L$ given by the NASA/IPAC Extragalactic Database\footnote{https://ned.ipac.caltech.edu/}.
We set the start time of the outburst as between the first data point in the rising phase and the previous data point, and calculated the $t_{\rm rise}$ as the time difference between the start and the peak time in the W2 band.
If the rising phase starts at the first data point of NEOWISE-R, or the outburst peaks at the last data point of the LC, we set the $t_{\rm rise}$ calculated above as the lower limit.
In addition, as can be seen from Figure~\ref{fig:example_ATs}, TDE AT 2017gge and ANT AT 2018bcb peaks at the first data point of the outburst state, and hence, we could only calculate the upper limit of $t_{\rm rise}$ for them.

\begin{figure}
\centering
 \includegraphics[scale=0.66]{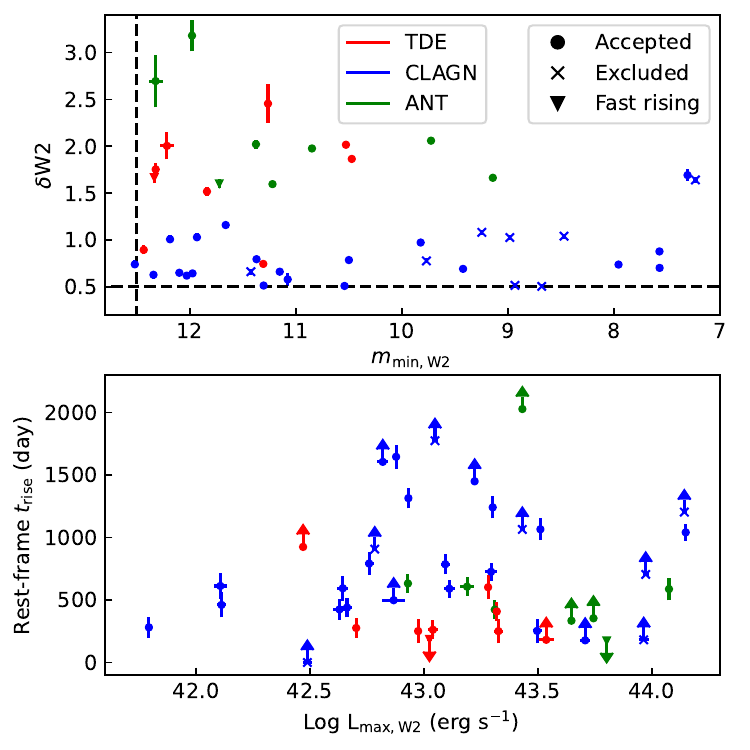}
 \caption{
  MIR properties of the final sample (dots) and the excluded CLAGNs (crosses).
  The upper panel shows $m_{\rm min,W2}$ and $\delta{\rm W2}$, and the lower panel shows $L_{\rm max,W2}$ and $t_{\rm rise}$.
  TDEs, ANTs and CLAGNs are shown in red, green and blue symbols.
  The two fast-rising objects (AT 2017gge and AT 2018bcb) are marked with inverted triangles.
  Note that one excluded CLAGN, NGC 4395, has a luminosity of $L_{\rm max,W2}=10^{39.4}$ erg s$^{-1}$ and is outside the scope of the lower panel.
  }
\label{fig:mirproperty_opt}
\end{figure}

We show the distribution of $m_{\rm min,W2}$ and $\delta{\rm W2}$ in Figure~\ref{fig:mirproperty_opt}(a).
CLAGNs generally cause smaller amplitude of variability with $\delta{\rm W2}<1.7$ magnitude, while TDEs and ANTs can cause variation with $\delta{\rm W2}$ up to $\sim3$ magnitude.
We also show the distribution of $L_{\rm max,W2}$ and $t_{\rm rise}$ in Figure~\ref{fig:mirproperty_opt}(b).
The MIR luminosities $L_{\rm max,W2}$ of the sample are mainly in the range of $\sim10^{42}-10^{44}$ erg s$^{-1}$.
The TDEs generally have lower MIR luminosities than ANTs as the median values of $L_{\rm max,W2}$ are $\sim10^{43}$ and $\sim10^{43.6}$ erg s$^{-1}$, respectively.
The rise time of the sample is distributed over a wide range between $<200$ days and $>2000$ days.
TDEs and ANTs generally have shorter rise time than CLAGNs, as among the 16 objects whose $t_{\rm rise}$ is $>2$ years, only one is TDE, and one is ANT, and the rest are CLAGNs.

The eight CLAGNs excluded in section \ref{sec:opt_rising} have similar properties to other CLAGNs, so excluding them would not change the overall properties of the sample.
This was reasonable because they were excluded due to the inappropriate start time for a detailed study of the rising phase.

\section{Analysis of MIR color variation and results} \label{sec:mircolor}

This section studied the MIR color variation of the optically selected outbursts in the rising phase.
We did not calculate the dust temperature as some previous works had done \citep[e.g.,][]{Jiang2017,Reynolds2022}.
This is because the dust temperature depends heavily on the spectral energy distribution (SED), and its relationship with the directly observed quantity is complicated, making its error difficult to analyze.
Instead, we used K-corrected MIR color $W1-W2$, i.e. color in the rest frame, as a probe of dust temperature.
We describe how we applied the K-correction and estimated the systematic error introduced by the K-correction in Appendix~\ref{appendix:kcorr}.

We calculated the color of the excess flux after subtracting the quiescent state flux.
We did not use the total flux because it contains emission from the host galaxy, which is unrelated to the outburst.
For outbursts occurring in AGNs, like CLAGNs and ANTs, the echoing dust is likely the AGN dusty torus, and hence the color we measured may present the change in dust temperature, rather than the temperature itself.
We will analyze this effect in section \ref{sec:simu_agn}.

\subsection{Time variation of MIR color in the rising phase} \label{sec:mircolor_cvr_ic}

\begin{figure}
\centering
 \includegraphics[scale=0.66]{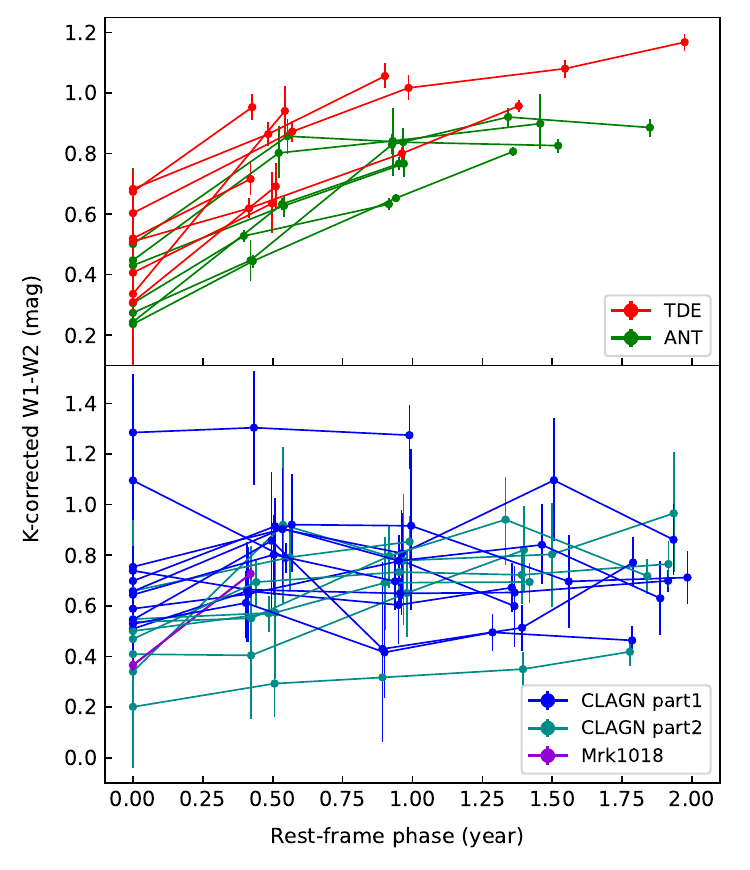}
 \caption{
  Time variation of MIR color in the rising phase.
  The upper panel shows TDEs (red) and ANTs (green), and the lower panel shows CLAGNs.
  For clarity, we split the CLAGN sample into two parts, one with CVR$<0.1$ (blue) and the other with CVR$>0.1$ (dark cyan).
  We labelled CLAGN Mrk 1018 with special color variation using purple.
  }
\label{fig:color_all}
\end{figure}

We calculated the K-corrected MIR color in the rising phase for all 37 objects.
In Figure~\ref{fig:color_all}, we show the variation of MIR color over time, grouped by different types.
For objects with $t_{\rm rise}>2$ years, we only show variation in the first two years.

The MIR color of TDEs and ANTs generally turns red (increases) quickly during the rising phase, while that of CLAGNs shows no significant trend, either unchanged or turning red slowly.
In addition, the color of ANTs is generally bluer than TDEs at the same phase.

To quantify these differences, we fit the time variation of color using a linear function and adopted the slope as color variation rate (CVR), and calculated its error using the Monte Carlo method.
We also adopted the color of the first point in the rising phase as the initial color (IC).
Note that AT 2017gge and AT 2018bcb have only one data point in the rising phase, so CVR could not be calculated.
We list the CVRs for the remaining 35 objects and the ICs for all 37 objects in Table~\ref{tab:information_CVR_IC}.

\begin{figure}
\centering
 \includegraphics[scale=0.66]{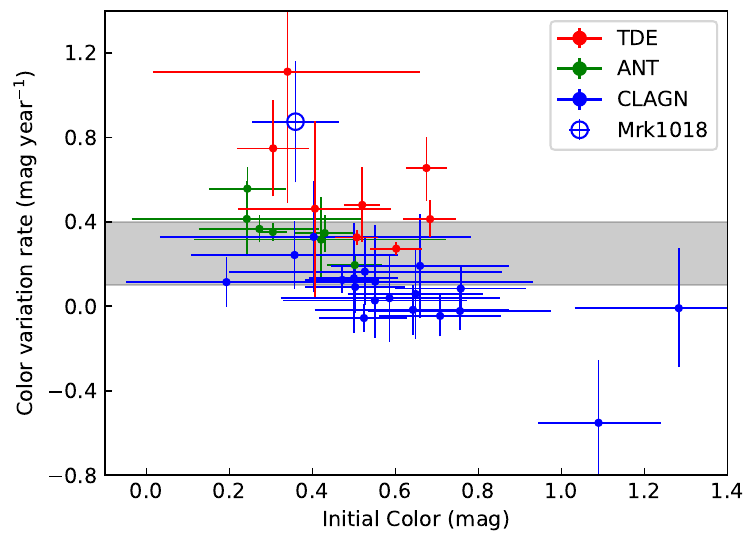}
 \caption{
  The distribution of ICs and CVRs of the optically selected sample.
  Red, green and blue represent TDE, ANT, and CLAGN, respectively.
  We specifically marked Mrk 1018, whose type is uncertain.
  We show the CVR thresholds used to distinguish TDE, ANT and CLAGN in a grey shade.}
\label{fig:cvr_ic_opt}
\end{figure}

As seen from Figure~\ref{fig:cvr_ic_opt}, TDEs and ANTs have overall larger CVRs than CLAGNs.
The CVRs of 8 TDEs are between 0.27 and 1.11 mag yr$^{-1}$ with a median of 0.47, and those of 7 ANTs are between 0.20 and 0.56 mag yr$^{-1}$ with a median of 0.35, and those of 20 CLAGNs are between $-0.55$ and 0.88 mag yr$^{-1}$ with a median of 0.09.
We checked the statistical significance of the difference in CVR distributions across TDEs, ANTs and CLAGNs using the K-S test.
The difference between TDEs and CLAGNs (ANTs and CLAGNs) is significant as the probability corresponding to the K-S statistic is $2.9\times10^{-5}$ ($3.1\times10^{-4}$).
At the same time, there is no significant difference between TDEs and ANTs.
In addition, the median IC of 8 ANTs (including AT 2018bcb) is 0.36 mag, smaller than 0.52 mag, the median of 9 TDEs (including AT 2017gge), and the difference between IC distributions is also significant according to the K-S test as the probability is 0.020.

The above analyses do not take into account the influence of measurement errors.
In general, for large enough sample sizes, measurement errors would cause the distribution of observed values to be more diffuse than that of the true values, which weakens the statistical significance of the difference between two distributions.
Therefore, the difference in CVR/IC distributions cannot be explained by measurement errors.

Among the 20 CLAGNs, Mrk 1018 has a CVR of $0.87\pm0.29$ mag yr$^{-1}$, significantly higher than other CLAGNs and similar to TDEs and ANTs.
Despite that Mrk 1018 is a famous CLAGN with long-term, large-amplitude variability along with state transition \citep[e.g.,][]{Cohen1986_Mrk1018,Husemann2016_Mrk1018}, it met our criteria because of its giant outburst in 2020 \citep{Brogan2023_Mrk1018,Lu2025_Mrk1018}.
We analyzed the optical LCs of Mrk 1018 during this outburst and present the results in Appendix~\ref{appendix:mrk1018}.
In brief, the lack of data makes it difficult to distinguish whether it is closer to typical CLAGN outbursts or ANTs like AT 2017bgt.
Therefore, we removed Mrk 1018 in the following analysis of the MIR colors of the sample because its type is uncertain.

\subsection{Distinguishing TDEs/ANTs from CLAGNs} \label{sec:mircolor_cvr_threshold}

Since TDEs/ANTs generally have larger CVRs than CLAGNs, we considered distinguishing TDE/ANT from CLAGN using the maximum CVR of CLAGNs and the minimum CVR of TDE+ANTs.
In the current sample, the maximum of the 19 CLAGNs (excluding Mrk 1018) is 0.33 mag yr$^{-1}$, while the minimum of the 15 TDE+ANTs is 0.20 mag yr$^{-1}$.
However, these values have two uncertainties, one from sampling and the other from measurement error.

We first considered the impact of these uncertainties on the maximum of CLAGNs.
The sampling leads to underestimation of the maximum, while measurement error leads to overestimation when the sample size is large enough.
The two uncertainties lead to opposite effects, whose strengths are related to the sample size, the intrinsic scatter of CLAGNs' CVRs, and the level of errors.
To quantitatively assess which effect is more substantial, we ran a simulation.
In the simulation, we assumed that the current observed values were true and evaluated the effects of sampling and measurement error using bootstrap and Monto Carlo methods, respectively.
Simulations show that the fake sample overestimated the maximum 97.5\% of the time, indicating that the overestimation caused by measurement error dominates the underestimation caused by sampling.
According to the simulation, the probability of the fake sample overestimating the maximum or underestimating by $<0.07$ mag yr$^{-1}$ is as high as 99.3\%.
Therefore, we considered it safe to set the upper limit of CLAGNs' CVR at 0.4 mag yr$^{-1}$.

We then analyzed the minimum CVR of TDE+ANTs similarly.
We found that the underestimation of the minimum caused by measurement error dominates the overestimation caused by sampling.
Similarly, we adopted a safe lower limit of TDE+ANTs' CVR at 0.1 mag yr$^{-1}$, which underestimates the true value by $>99.9\%$ according to the simulations.

We finally divided Figure~\ref{fig:cvr_ic_opt} into three regions based on the two thresholds of CVRs (grey shades):
region U with ${\rm CVR}>0.4$ mag yr$^{-1}$ representing a high probability of TDE or ANT, region L with ${\rm CVR}<0.1$ mag yr$^{-1}$ representing a high probability of CLAGN, and region M with $0.1\leq{\rm CVR}\leq0.4$ mag yr$^{-1}$, where the type of the object is difficult to determine.

The CVR could not be calculated for AT 2017gge and AT 2018bcb because they rise too fast relative to the cadence of WISE.
Both objects are TDE or ANT, consistent with that TDEs/ANTs generally have shorter $t_{\rm rise}$ than CLAGNs (section \ref{sec:opt_mir_property}).
This means an object rising too fast to calculate CVR is more likely to be TDE/ANT than CLAGN.
However, the confidence level of this judgement is not high due to the small sample sizes.

\section{Explanations for differences in MIR color variations} \label{sec:simulation}

In this section, we explore the possible explanations of the differences in MIR color variation between different types of outbursts with the help of simulations.
The main phenomena we need to explain are: 1. MIR color of TDEs and ANTs turns red faster than that of CLAGNs in the rising phase, as the CVRs of TDE+ANTs and CLAGNs are $\gtrsim0.2$ and $\lesssim0.3$ mag yr$^{-1}$, respectively; 2. ANTs are bluer than TDEs when they begin, as their median IC is smaller than TDEs.

We simulated the IR LCs of dust echoes of TDEs, ANTs and CLAGNs using a dust radiative transfer model developed by \citet[][hereafter Lu16]{Lu2016}.
The model was described in detail in section 3 of Lu16.
In brief, it calculates the IR reradiation of a dust spherical shell heated by a transient UV source in the galaxy center.
The LC of the transient source is described as $L_{\rm UV}(t)$.
The dust shell has an initial inner radius of $r_{\rm in}$ and an outer radius of $r_{\rm out}$.
The dust composition is a standard ISM mixture of 47\% graphite and 53\% silicate.
The dust grain has a radius of $a_0$, and is uniformly distributed with a number density of $n_d$.
This model has the advantage in that it calculates the decrease in the dust grain size caused by sublimation by using thermal equilibrium, thereby consistently obtaining how the inner radius of the dust structure increases over time.

For CLAGNs and ANTs, the echoing dust is likely the dusty torus in the AGN unified model \citep{Antonucci1993}.
In order to simulate the echo of torus dust, we developed Lu16's model as described in the appendix~\ref{appendix:simu_torus}.
In the torus dust case, the model has two more free parameters, including the half-opening angle and the inclination angle $i$.

We investigated the influence of each parameter on the MIR color variation using the control variable method.
We display the influence of different UV LCs in section \ref{sec:simu_uvlc}, the influence of different echoing dusts in section \ref{sec:simu_dust}, and finally discuss possible explanations of different CVR and IC in section \ref{sec:simu_explanation}.

\subsection{The difference in UV LCs} \label{sec:simu_uvlc}

\subsubsection{Influence of shape of UV LC} \label{sec:simu_uvshape}

The shapes of UV LCs of TDEs and ANTs generally differ from those of CLAGNs.
TDEs typically rise of $\sim10-50$ days, peak of 1--3 months, and fade of months to years \citep[e.g.,][]{vanVelzen2020_review,Yao2023}, and ANTs have similar UV LCs to TDEs.
However, the LCs of CLAGNs are not regular, and their rises are slower with rise time usually between months to years \citep[e.g.,][]{Ricci2023}.

To reflect this difference in the shape of UV LC, we tried four forms of UV LC.
The first is a TDE form where $L_{\rm UV}$ rises rapidly and fade slowly as $L\propto t^{-5/3}$:
\begin{equation}
L_{\rm UV}(t) =
    \begin{cases}
    0 & \text{ , $t<t_0$ } \\
    L_{\rm max} \left( 1 + \frac{t-t_0}{\tau_1} \right)^{-5/3} & \text{ , $t>t_0$, }
    \end{cases}
\end{equation}
where $t_0$ is the outburst time, $L_{\rm max}$ is the peak UV luminosity and $\tau_1$ describes the rate of fading.
The second is an EXP form where $L_{\rm UV}$ rises rapidly and fade exponentially:
\begin{equation}
L_{\rm UV}(t) =
    \begin{cases}
    0 & \text{ , $t<t_0$ } \\
    L_{\rm max} e^{-\frac{t-t_0}{\tau_2}} & \text{ , $t>t_0$, }
    \end{cases}
\end{equation}
The third is a LINEAR form where $L_{\rm UV}$ rises and falls linearly with similar rising and falling time scales, and the LC is described as:
\begin{equation}
L_{\rm UV}(t) =
    \begin{cases}
    L_{\rm max}\frac{t-t_0}{\tau_3} & \text{ , $ t_0 < t < t_0 + \tau_3 $ } \\
    L_{\rm max} \left( 2-\frac{t-t_0}{\tau_3} \right) & \text{ , $ t_0 + \tau_3 < t < t_0 + 2\tau_3 $ } \\
    0& \text{ , else, }
    \end{cases}
\end{equation}
And the fourth is a FLAT form expressed as:
\begin{equation}
L_{\rm UV}(t) =
    \begin{cases}
    L_{\rm max} & \text{ , $ t_0 < t < t_0 + \tau_4 $ } \\
    0& \text{ , else, }
    \end{cases}
\end{equation}
We referred to the duration $\tau$ of an outburst as the ratio of the total energy to the peak luminosity:
\begin{equation}
\tau = \frac{ \int_{-\infty}^{+\infty}L(t){\rm d}t }{ L_{\rm max} }
\end{equation}
For the four forms, $\tau$ equals to $\frac{2}{3}\tau_1$, $\tau_2$, $\tau_3$ and $\tau_4$, respectively.

We estimated the typical $L_{\rm max}$ and $\tau$ for TDEs and CLAGNs.
The typical $L_{\rm max}$ of TDEs are $10^{43.0}-10^{45.4}$ erg s$^{-1}$ \citep[e.g.,][]{Yao2023}.
We calculated the durations $\tau$ of ZTF TDEs to be $\sim40-200$ days using the LC models of \citet{vanVelzen2021}.
On the other hand, CLAGNs' $L_{\rm max}$ and $\tau$ range widely.
Assuming that $L_{\rm UV}$ is $\sim$10 times $L_{\rm W2}$ according to typical AGNs' SEDs, the CLAGNs in this work have $L_{\rm max}$ in the range of $\sim10^{43}-10^{45}$ erg s$^{-1}$.
Most CLAGNs last for years to decades \citep[e.g.,][]{Guo2024}, while some last for only a few months, as seen in the 2018 outburst of NGC 1566 \citep{Oknyansky2019}.
Thus, in the simulations, we set the default values of $L_{\rm max}=10^{44}$ erg s$^{-1}$ and $\tau=10^7$ s.
%The $L_{\rm max}$ of MIR bright TDEs in our sample are between $10^{43.4}$ and $10^{45.7}$ erg s$^{-1}$ \citep{vanVelzen2021,Onori2022,Wang2024_AT2018gn,Earl2025,Zhang2025}

We assumed a grain size $a_0=0.1$ $\mu$m.
We set the default value of initial inner radius $r_{\rm in}$ as $0.6 r_{\rm sub}(L_{\rm max})$, where $r_{\rm sub}$ is the sublimation radius corresponding to $L_{\rm max}$, and is calculated as \citep[e.g.,][]{Peterson1997}:
\begin{equation}
r_{\rm sub}(L_{\rm max}) = 0.4 \left( \frac{ L_{\rm max} }{ 10^{45}\ {\rm erg}\ {\rm s}^{-1} } \right) ^{1/2} {\rm pc},
\end{equation}
The dust radiative transfer model we used can calculate a new inner radius by taking the dust sublimation into account, and hence the value of $r_{\rm in}$ has little influence on IR LCs.
The outer radius $r_{\rm out}$ and grain density $n_d$ affect the UV optical depths of the dust shell as:
\begin{equation}
\begin{aligned}
\tau_{\rm UV} &\approx (r_{\rm out} - r_{\rm sub})\pi a_0^2 n_d Q_{\rm UV} \\
&\approx 20 \left( \frac{ r_{\rm out} - r_{\rm sub} }{\rm pc} \right) \left( \frac{n_d}{10^{-8}\ {\rm cm}^{-3}}  \right),
\end{aligned}
\end{equation}
where $Q_{\rm UV}=2$ is the UV extinction coefficient factor assuming equal contributions from absorption and scattering.
We estimated the typical $\tau_{\rm UV}$ for dusty TDEs as follows.
On one hand, dusty TDEs have $L_{\rm IR}$ close to $L_{\rm UV}$, requiring that $\tau_{\rm UV}\gtrsim1$.
On the other hand, they show strong silicate emission features \citep{Masterson2025}, indicating that the dust is optically thin in the MIR band with $\tau_{\rm MIR}<1$.
For a mixture of graphite and silicate dust with a grain size of 0.1 $\mu$m, the MIR absorption coefficient factor at wavelengths of silicate features is $\approx0.06$, and then $\tau_{\rm MIR} \approx 0.03 \tau_{\rm UV}$.
Combining these, $\tau_{\rm UV}$ in TDEs should be in the range of $\sim1-30$.
Thus, we set default values of $r_{\rm out}$ and $n_d$ as 1 pc and $10^{-8}$ cm$^{-3}$, respectively, yielding a $\tau_{\rm UV}$ of $\sim20$, close to those of dusty TDEs and AGN tori \citep[several tens, e.g.,][]{Nenkova2008}.

\begin{figure}
\centering
 \includegraphics[scale=0.66]{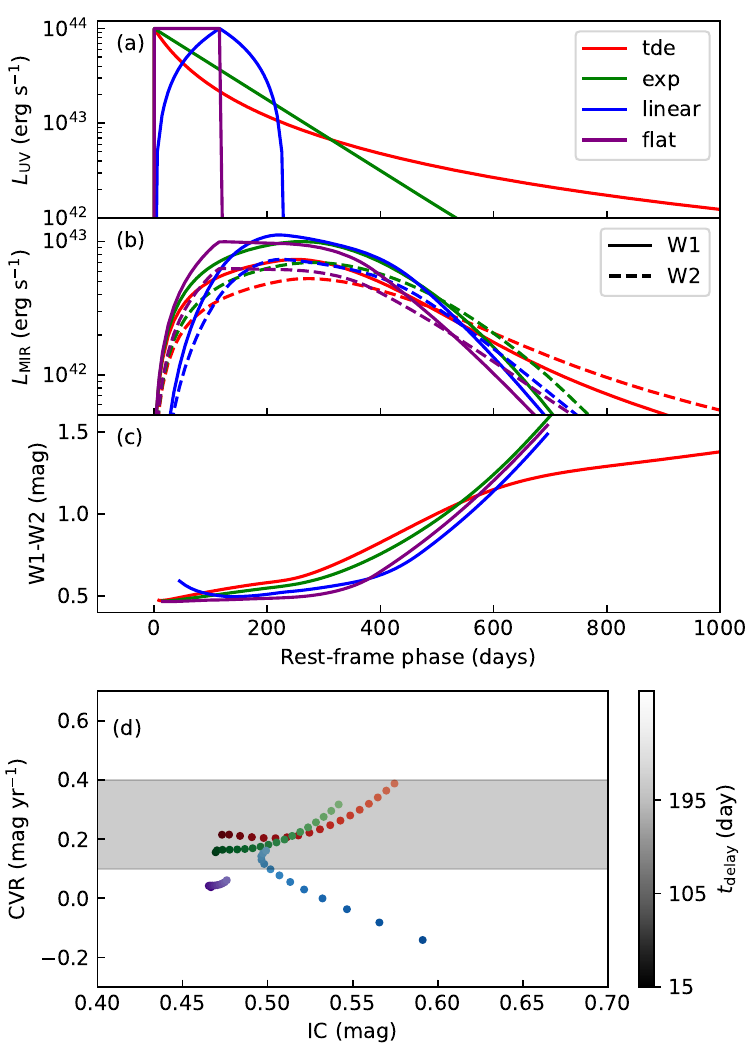}
 \caption{
  \textbf{(a)}: The assumed UV LCs with TDE (red), EXP (green), LINEAR (blue) and FLAT (purple) forms.
  \textbf{(b)}: The simulated MIR LCs in the W1 (solid) and W2 (dashed) bands.
  \textbf{(c)}: The inferred W1-W2 color curves.
  \textbf{(d)}: The IC and CVR calculated using the MIR color data points sampled every 180 days.
  Values assuming different sampling times are differentiated by color saturations.
  }
\label{fig:simu_uvlc_shape}
\end{figure}

We simulated the MIR LCs and calculated CVRs and ICs for different shapes of UV LC.
As shown in Figure~\ref{fig:simu_uvlc_shape}, the MIR color curves in the rising phases are different.
For the TDE and EXP forms, initially, the MIR color is relatively blue and becomes red continuously in the rising and fading phases.
While for the LINEAR and FLAT forms, the MIR color turns slightly blue or keeps steady with the rapid rise of luminosity in the early time, and then keeps a blue color near the peak, and finally turns red as it fades.
We sampled the simulated MIR color curve at 180-day intervals and calculated IC and CVR in the rising phase using the sampled data.
IC and CVR are not definite values but in a range because of the uncertain sampling time, expressed as the time delay $t_{\rm delay}$ between the first sampling and the outburst time.
We found that the form of LC has little effect on the IC, but affected the CVR: CVR is 0.2--0.4, 0.15--0.35, $-0.1$--0.2 and $\sim0.05$ mag yr$^{-1}$ for TDE, EXP, LINEAR and FLAT forms.
We also tried other forms of UV LC with more complicated models.
By summarizing these results, we found that to produce a large CVR of $\gtrsim0.3$ mag yr$^{-1}$, the UV LC needs to contain a rapid rise, a short peak, and a long tail.
The reason is that the echoes of the peak luminosity and the long tail have blue and red MIR colors, respectively, and in the rising phase, the fraction of the echo of the long tail is increasing, causing the MIR color to turn red continuously.

\subsubsection{Influence of underlying AGN} \label{sec:simu_agn}

\begin{figure*}
\centering
 \includegraphics[scale=0.85]{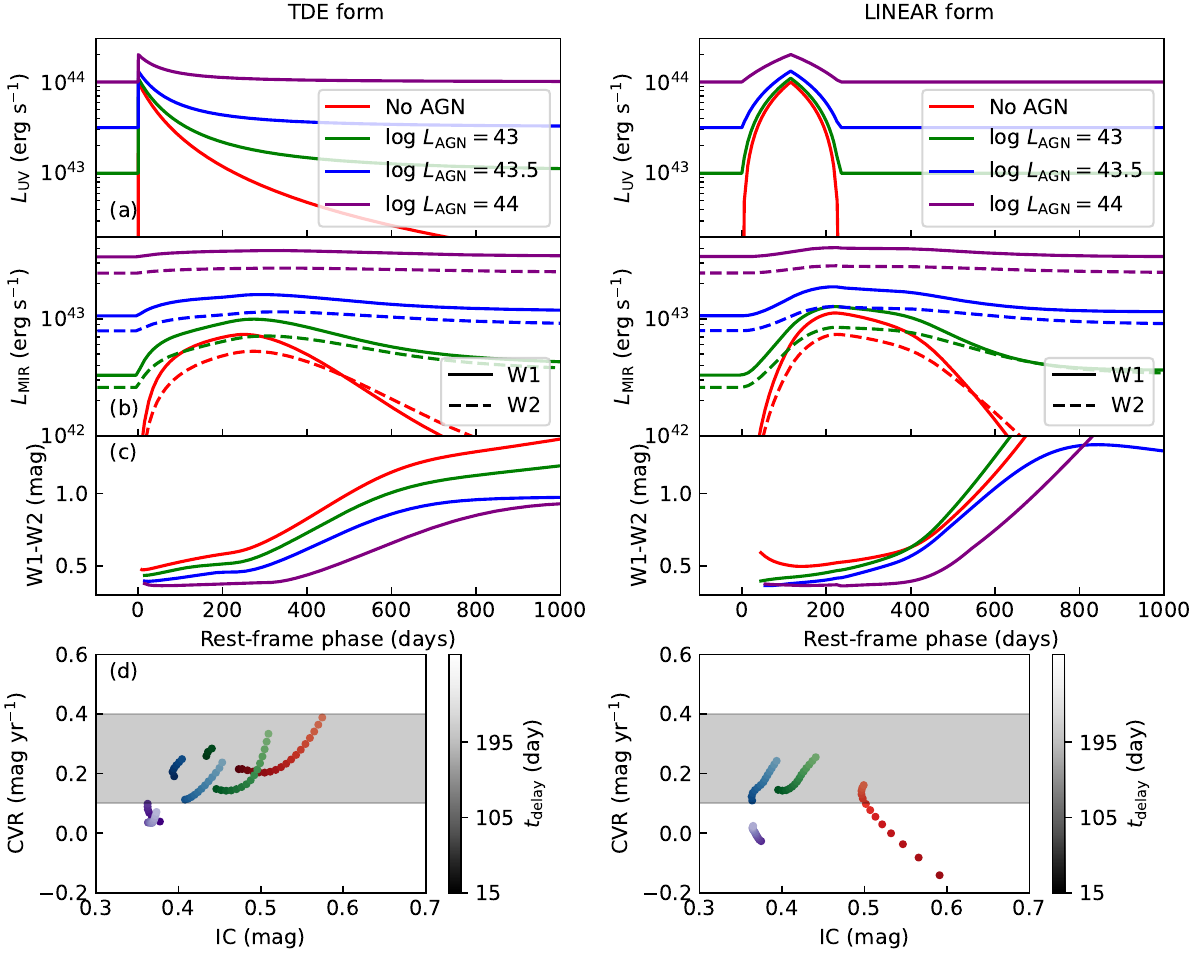}
 \caption{
  Simulations exploring the influence by underlying AGN.
  The meanings of panels (a) to (d) are the same as Figure~\ref{fig:simu_uvlc_shape}.
  Left panels show simulations assuming TDE-form UV LCs, and right panels show those assuming LINEAR-form UV LCs.}
\label{fig:simu_lagn}
\end{figure*}

ANTs and CLAGNs occur in AGNs with long-lived IR emission from the dusty torus.
The underlying AGN causes multiple simultaneous effects, and we explored these effects using simulations.
In the simulations, we set the UV light curve of the illuminating source as:
\begin{equation}
L_{\rm UV}(t) = L_{\rm flare}(t) + L_{\rm AGN},
\end{equation}
where $L_{\rm flare}(t)$ is the UV light curve after subtracting the underlying AGN, and $L_{\rm AGN}$ is the UV luminosity of the underlying AGN.
For $L_{\rm flare}(t)$, we set TDE and LINEAR form UV LCs for ANTs and CLAGNs, respectively, and set $L_{\rm max}$ of $10^{44}$ erg s$^{-1}$ and $\tau$ of $10^7$ s.
We set $L_{\rm AGN}$ to be in the range of $10^{43}$--$10^{44}$ erg s$^{-1}$, corresponding to amplitudes of variability of 0.7--2.6 magnitude, roughly consistent with the observed $\delta W2$ of 0.5--3.2 magnitude.
Note that we subtracted the quiescent fluxes when calculating the MIR color in Section 3, so in the simulation, we also calculated the MIR color after subtraction the IR luminosities of the underlying AGN (the values at $t<=t_0$).
The results are shown in Figure~\ref{fig:simu_lagn}, and we also show those with no underlying AGN for a comparison.

% \textbf{which can be estimated by assuming a black hole mass of $10^7$--$10^8 M_{\bigodot}$ and using the CLAGN Eddington ratio of \cite{Panda2024}.}

We found that for the TDE form UV LC, as the ratio of AGN luminosity to the peak outburst luminosity is larger, indicating a smaller amplitude of variability, both the CVR and IC are smaller and deviate more from the values without AGN.
While for the LINEAR form UV LC, underlying AGNs little affect the median value of CVR.
Thus, in order to produce a large CVR, there must be no AGN or a relatively small AGN contribution.

ANTs generally have large amplitudes of variability: optical ANTs can be brightened by $\gtrsim2-3$ magnitudes in the UV bands at peak.
Thus, the AGN contribution results in ANTs having smaller CVRs than TDEs, but the difference between the two is no more than $\sim$0.1 mag yr$^{-1}$.
On the other hand, most CLAGNs have smaller amplitudes of UV variability of $\lesssim1-2$ magnitudes.
Thus, even if a CLAGN happened to have a UV LC similar to those of TDEs, its CVR would not be large, in the case that it does not vary sufficiently.

\subsubsection{Influence of peak UV luminosity and duration} \label{sec:simu_lmax_tau}

In subsequent simulations, we will focus on two cases: a TDE case with a TDE form $L_{\rm UV}(t)$ and no underlying AGN, and a CLAGN case with a LINEAR form $L_{\rm flare}(t)$ and an underlying AGN with $L_{\rm AGN}=L_{\rm max}$.

\begin{figure*}
\centering
 \includegraphics[scale=0.85]{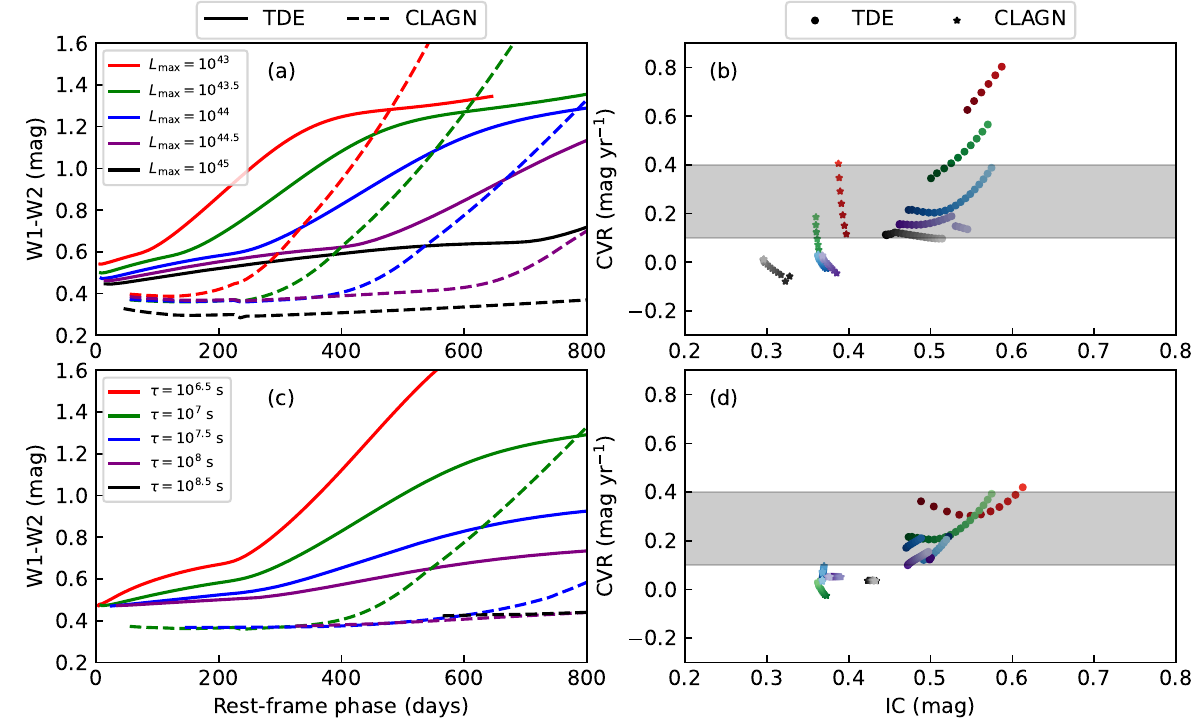}
 \caption{
  Simulations exploring the influence by different $L_{\rm max}$ and $\tau$.
  Panels (a) and (c) show the MIR color curves (solid and dashed for the TDE and CLAGN cases, respectively), and (b) and (d) show the inferred ICs and CVRs using the sampled data (circles and stars for the TDE and CLAGN cases, respectively).
  Different colors indicate different $L_{\rm max}$ for panels (a) and (b), and indicate different $\tau$ for panels (c) and (d).}
\label{fig:simu_lmax_tau}
\end{figure*}

We studied the influence of different UV peak luminosities $L_{\rm max}$ in the range of $10^{43}-10^{45}$ erg s$^{-1}$ (Figure~\ref{fig:simu_lmax_tau}(a), (b)).
For both the TDE and CLAGN cases, CVR decreases with the increase of $L_{\rm max}$, while IC is little affected.
The $L_{\rm max}$ of TDEs ($10^{43}-10^{45.4}$ erg s$^{-1}$) are similar to that of CLAGNs in our sample ($\sim10^{43}-10^{45}$ erg s$^{-1}$).
Thus, the difference in $L_{\rm max}$ can hardly explain that TDEs have higher CVRs than CLAGNs.
In addition, the dispersion of $L_{\rm max}$ of about two orders of magnitude may explain the internal CVR dispersion among TDEs, ANTs or CLAGNs.

We then studied the influence of different durations $\tau$ in the range of $10^{6.5}-10^{8}$ s for the TDE case and $10^{7}-10^{8.5}$ s for the CLAGN case (Figure~\ref{fig:simu_lmax_tau}(c), (d)).
For the TDE case, CVR in the rising phase decreases with the increase of $\tau$.
When $\tau$ increases to $\sim10^{7.5}-10^8$ s ($\sim1-3$ years), the CVR drops to values similar to those in CLAGNs.
Fortunately, the durations of most TDEs and ANTs ($\sim40-200$ days) are not that long, and thus the results obtained by assuming $\tau=10^7$ s are not affected.
While for the CLAGN form, CVR is little affected by $\tau$.

\subsection{Difference in echoing dust} \label{sec:simu_dust}

\subsubsection{Influence by dust properties} \label{sec:simu_dust_property}

The chemical composition and grain size of the dust influence the MIR color.
As shown in Figure~\ref{fig:color_kcorr}(a), graphite dust is bluer than silicate dust at the same temperature.
Moreover, the sublimation temperature of graphite ($\sim1900$ K) is higher than that of silicate ($\sim1500$ K), resulting in a much higher MIR color of graphite ($\sim0.2$ mag) than silicate ($\sim-0.3$ mag) at around sublimation temperature.
Thus, the difference in dust composition can affect the IC: a higher fraction of graphite results in a bluer MIR color (smaller IC).
Dust grains with different radii with the same composition show a small difference in color of 0.1--0.2 mag at the same temperature.
Thus, the grain size may slightly affect the IC.

To explain that ANTs have bluer MIR colors than TDEs, it is required to assume that the echoing dust of ANTs, which is likely the inner region of AGN torus, has a higher graphite fraction and a larger grain size than the echoing dust of TDEs.
However, although we have some knowledge of the dust properties in AGNs, those in TDEs are poorly constrained by observations.
Future timely follow-up spectroscopies of dusty TDEs may constrain the dust properties, check whether they differ from AGN tori, and explore if they can explain the observed IC difference between TDEs and ANTs.

Does the difference in dust properties affect CVR?
Answering this question requires considering that dust composition may be dependent on locations.
Assuming that the initial dust is a mixture of graphite and silicate, the innermost region of the final dust structure would have a high graphite fraction because the sublimation radius of the graphite is larger than that of the silicate.
This scenario was supported by observations, as \citet{Masterson2025} found that the MIR spectrum of a dusty TDE, WTP14adbjsh, can be explained with a sum of a close graphite dust component and a distant mixed dust component.
After taking this effect into account, the color difference between the inner region (arrives earlier) and the outer region (arrives later) would be larger, resulting in a larger CVR.
However, this effect may work in both TDE and CLAGN, so it might not explain the difference in CVR between the two samples.

\subsubsection{Influence by inner and outer radii} \label{sec:simu_inner_outer}

For TDEs and AGNs, the inner boundary of the echoing dust is generally determined by dust sublimation.
In this case, it would be better to describe the actual inner radius with $r_{\rm sub}$, which is determined by $L_{\rm max}$ and is not an independent parameter.
Thus, the setting of initial inner radius $r_{\rm in}$ in the simulation would have little influence as long as $r_{\rm in}<r_{\rm sub}$.
This was confirmed by our simulations where we varied $r_{\rm in}$.

\begin{figure*}
\centering
 \includegraphics[scale=0.85]{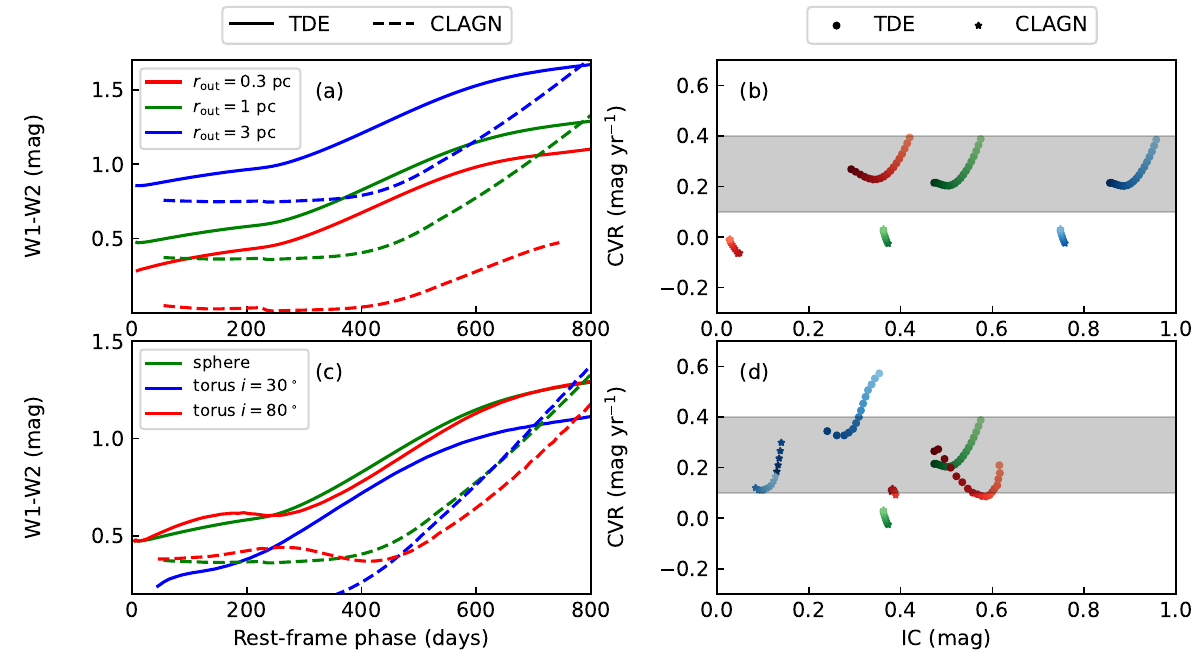}
 \caption{
  Simulations exploring the influence of the shape of dust structure.
  Panels (a) and (c) show the MIR color curves (solid and dashed for the TDE and CLAGN cases, respectively), and (b) and (d) show the inferred ICs and CVRs using the sampled data (circles and stars for the TDE and CLAGN cases, respectively).
  Different colors indicate different $r_{\rm out}$ for panels (a) and (b), and different geometry for panels (c) and (d).}
\label{fig:simu_dust_shape}
\end{figure*}

%We first studied the influence of the initial inner radius $r_{\rm in}$ on CVR and IC (Figure~\ref{fig:simu_dust_shape}(a), (b)).
%The results show that $r_{\rm in}$ has little effect on the MIR color variation when it is much smaller than $r_{\rm sub}$.
%However, when $r_{\rm in}$ is close to or greater than $r_{\rm sub}$, the increase of $r_{\rm in}$ leads to the increase of IC for both TDE and CLAGN cases, and leads to the decrease of CVR for the TDE case.
%The model can calculate a new inner radius by taking the dust sublimation into account, and hence the value of $r_{\rm in}$ has little influence on the IR LCs when it is less than $0.8 r_{\rm sub}$.
%In simulations of the main text, we set the default value of initial inner radius $r_{\rm in}$ as $0.6 r_{\rm sub}(L_{\rm max})$.

%Some CLAGNs have experienced multiple outbursts in their history, for example NGC 1566 \citep{Alloin1986,Oknyansky2019}.
%If a stronger outburst than the current one occurred in history, the $r_{\rm in}$ of the echoing dust may be larger than the $r_{\rm sub}$ corresponding to the peak luminosity of the current outburst.
%This may explain why NGC 1566 has the smallest CVR ($-0.55\pm0.29$ mag yr$^{-1}$) in our sample and a relatively large IC ($1.09\pm0.15$ mag).

We then studied the influence of the outer radius $r_{\rm out}$ in the range of 0.3--3 pc.
We show the simulation results in Figure~\ref{fig:simu_dust_shape}(a), (b).
The $r_{\rm out}$ has little effect on CVR, but significantly affects IC.
This is because the IR emission in the rising phase mainly comes from the inner region of the dust structure, and the outer region only affects the optical depth of the emergent IR emission.
The increase of $r_{\rm out}$ leads to the increase of $\tau_{\rm MIR}$, thus causing the MIR color to turn red.
However, this reddening hardly changes over time, and thus has little effect on CVR.

Can the difference in MIR emergent optical depth explain why ANTs have smaller (bluer) ICs than TDEs?
As we have demonstrated in section~\ref{sec:simu_uvshape}, the echoing dust in TDEs has lower UV optical depth than the AGN dusty torus.
However, ANTs have bluer MIR colors than TDEs, suggesting smaller MIR emergent optical depth.
These two are contradictory under the assumption of spherical shell.
We will consider other dust geometries in the following subsection.

\subsubsection{Spherical shell or torus} \label{sec:simu_sphere_torus}

We studied the influence of CVR and IC by dust geometry (spherical shell or torus) and the inclination angle of the torus.
We ran simulations for both types of geometry.
For the torus geometry, we considered two inclinations, $30^\circ$ for face-on torus and $80^\circ$ for edge-on torus.
For both cases the half-opening angle was set to be $40^\circ$.
The results are shown in Figure~\ref{fig:simu_dust_shape}(c), (d).

Face-on torus results in much smaller (bluer) IC than edge-on torus and spherical shell.
The reason is that the initial MIR emission, mainly from the innermost dust, emerges with a much smaller optical depth.
The difference in IC is positively correlated with the radial optical depth of the spherical shell or edge-on torus.
In our simulations where we assumed $\tau_{\rm UV}\sim20$, the IC difference is approximately 0.25 magnitude.

The ANTs in our sample are selected in the optical band, and hence, their tori are more likely to be face-on.
While for TDEs, the dust geometry might be better described by a spherical shell since there is no clear evidence for the anisotropy of nuclear dust structure in inactive galaxies.
Therefore, the difference in the dust geometry can explain why ANTs have smaller ICs than TDEs.

In addition, face-on torus results in a slightly larger CVR than the spherical shell with a difference of $\sim0.1-0.2$ mag yr$^{-1}$.
The CVR difference may be because the echo produced by the peak luminosity is bluer due to smaller emergent optical depth, and has a larger color difference with the echo from outer region with red color.

\subsection{Promising explanations} \label{sec:simu_explanation}

According to simulations in section \ref{sec:simu_uvlc}, in order to produce a large CVR, the UV LC must meet two conditions, one is that the LC of the outburst must be characterized by ``rapid rise, short peak and long tail'', and the other is that there is no or relatively weak underlying AGN.
The CVR difference between TDEs/ANTs and CLAGNs could be related to these two factors, while could hardly be associated with the differences in peak luminosities or durations.
In addition, we did not prefer explanations involving differences in dust structure or in dust properties, because ANTs and CLAGNs both occur in AGNs and there is no clear reason why their dust environments are different.

We preferred to explain the smaller ICs in ANTs than in TDEs by differences in dust property or dust structure, because the two types of outbursts have similar UV LCs.
The difference in dust properties requires that the echoing dust of ANTs, i.e., the inner region of an AGN dusty torus, has a higher fraction of graphite and larger grain size than the echoing dust of TDEs.
Whereas the difference in dust geometry requires that the echoing dust in ANTs is a face-on torus, while the echoing dust in TDEs is a spherical shell.
If the difference in dust geometry is the dominant factor, we predicted that the IC difference between TDEs and ANTs in MIR samples would be smaller or even disappear, because dusty tori in MIR-selected ANTs should have larger inclination angles than optically selected ANTs.

In the simulations, we assumed a smooth dust distribution, but there is evidence that the actual AGN dusty torus is clumpy \citep[e.g.,][]{Nenkova2008}.
Fortunately, the differences between the SEDs predicted by clumpy and smooth dust models mainly exist in the NIR spectral index and the strength of silicate feature, and the differences are slight at 3.4 and 4.6 $\mu$m \citep{Feltre2012}.
Thus, the results we obtained through simulations may still hold when taking clumpy dust into account.

\section{Applying to MIR outbursts} \label{sec:mir_sample}

As we have developed a new method to distinguish TDEs/ANTs from CLAGNs, we applied it to samples of MIR outbursts, including the mid-infrared outbursts in nearby galaxies (MIRONG) sample of \citet{Jiang2021}, and the sample of \citet{Masterson2024}, which is part of the Wise Transient Pipeline (WTP) sample.

\subsection{MIRONG sample} \label{sec:mir_mirong}

The MIRONG sample contains 137 MIR outbursts.
We removed four supernovae and four objects with strong radio emissions that may be associated with a jet, all identified by \citet{Jiang2021}.
We also removed two objects that contain problematic data points.
We reduced the data for the remaining 127 objects and selected quiescent and outburst states and the rising phase as we did for the optically selected sample.
The only difference is that we did not require $m_{\rm min,W2}<12.5$ or $\delta{\rm W2}>0.5$ here.
In this process, we removed 12 objects whose rising phase began at the first data point of NEOWISE-R.
We also removed nine objects with only one data point in the rising phase.
Finally, there are 106 objects left for the following analysis.

\begin{figure}
\centering
 \includegraphics[scale=0.66]{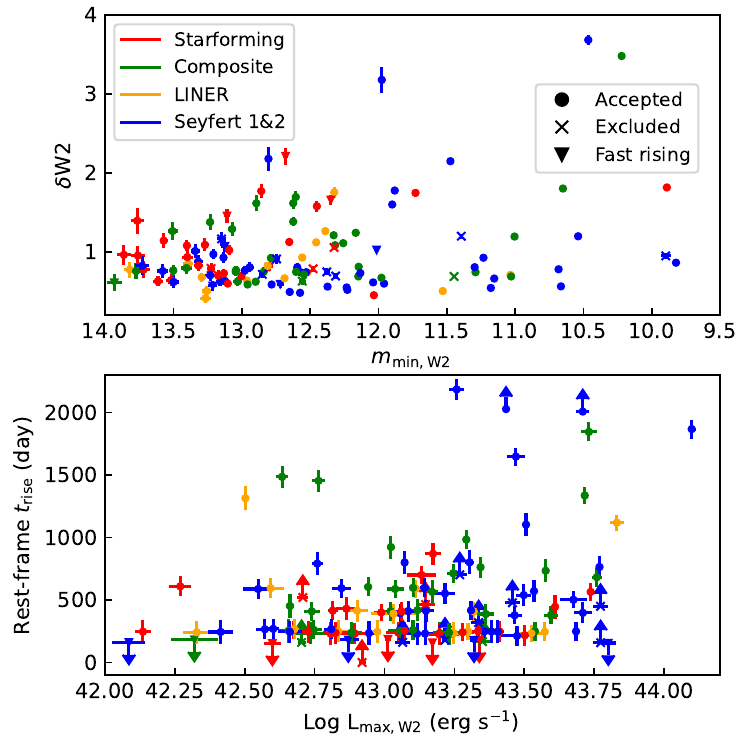}
 \caption{
  MIR properties of the MIRONG sample.
  The upper panel shows $m_{\rm min,W2}$ and $\delta{\rm W2}$, and the lower panel shows $L_{\rm max,W2}$ and $t_{\rm rise}$.
  Star-forming, Composite, LINER and Seyfert galaxies are shown in red, green, orange and blue symbols.
  Note that the objects in \citet{Jiang2021} with broad emission lines (Seyfert 1) and those with Seyfert-like narrow line ratios (Seyfert 2) were both treated as Seyfert galaxies in this work.
  The meanings of symbols (dot, cross and inverted triangle) are the same as Figure~\ref{fig:mirproperty_opt}.}
\label{fig:mirproperty_j21}
\end{figure}

The MIR properties of the 106 objects and the 21 objects removed when selecting the rising phase are displayed in Figure~\ref{fig:mirproperty_j21}.
The MIR luminosity $L_{\rm max,W2}$ of the sample is in the range of $\sim10^{42}-10^{44}$ erg s$^{-1}$, and the rise time $t_{\rm rise}$ is in the range of $<200$ and $>2000$ days.
These ranges are similar to those of the optically selected sample, and hence, this sample is suitable for the method distinguishing TDE/ANT and CLAGN developed in section 3.

\begin{figure*}
\centering
 \includegraphics[scale=0.88]{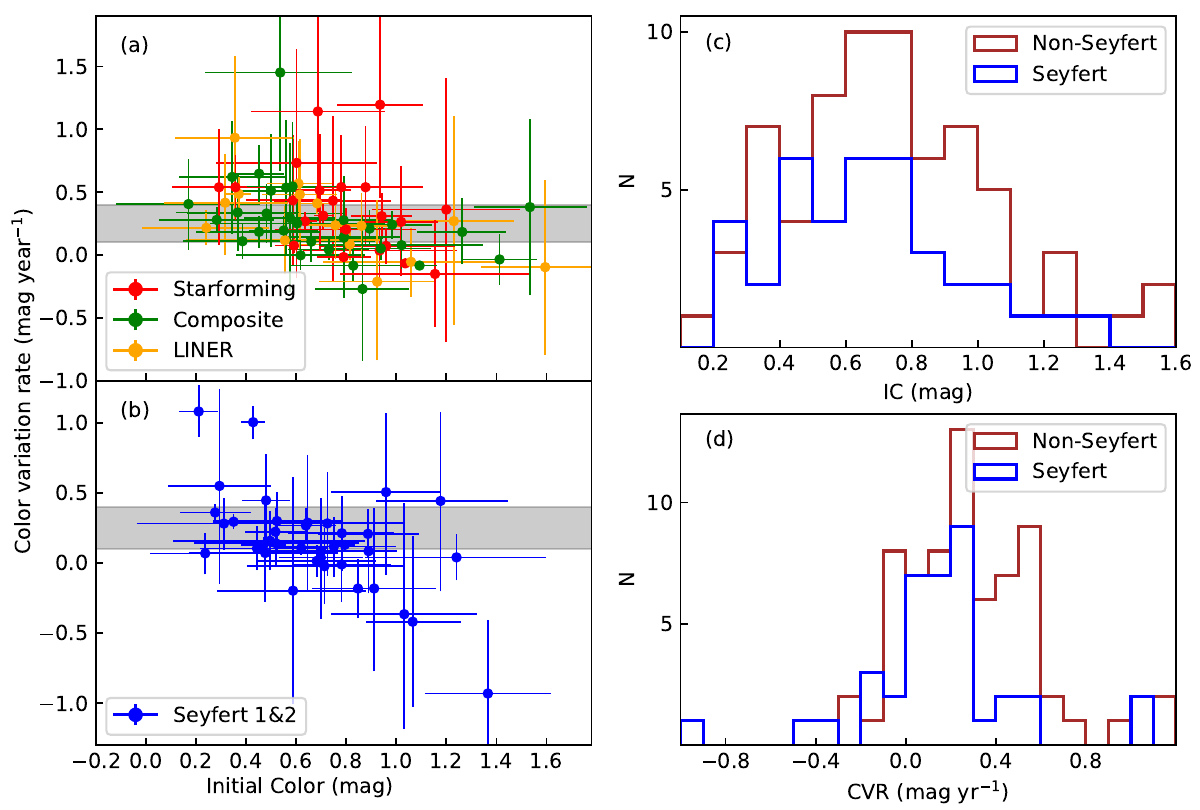}
 \caption{
  Distribution of ICs and CVRs of the MIRONG sample.
  Panel (a) shows star-forming (red), composite (green) and LINER (orange) galaxies and panel (b) shows Seyfert galaxies (blue).
  We compare the distributions of ICs for non-Seyfert and Seyfert galaxies in panel (c) and those of CVRs in panel (d).}
\label{fig:cvr_ic_j21}
\end{figure*}

We calculated the IC and CVR of the 106 MIR outbursts in the same way as we did for the optically selected sample.
Using a Monte Carlo method, we calculated the errors of IC and CVR, and also the probability that each object falls in the U, M, and L regions ($p_U$, $p_M$ and $p_L$) defined in section \ref{sec:mircolor_cvr_threshold}.
We list these results in Appendix \ref{appendix:param_mir}.

We show the distribution of ICs and CVRs in Figure~\ref{fig:cvr_ic_j21}, grouped into four subsamples according to the host galaxy type.
We made K-S tests to check whether the IC and CVR distributions are different among the subsamples.
The tests showed that there was no difference among star-forming ($N=22$), composite ($N=31$) and LINER ($N=15$) galaxies.
At the same time, there was a slight difference between CVRs of Seyfert ($N=38$) and non-Seyfert galaxies (the union of the other three, $N=68$) with a significance of 0.094 (Figure~\ref{fig:cvr_ic_j21}(d)).
The difference can also be seen from the median values: the median CVR of non-Seyfert galaxies is 0.27 mag yr$^{-1}$, which is between those of optically selected TDEs (0.47) and CLAGNs (0.09), while the median CVR of Seyfert galaxies is 0.13 mag yr$^{-1}$, close to that of CLAGNs.

\begin{table}
\centering
\caption{Numbers of MIR outbursts falling into the three regions.}
\begin{tabular}{ccccccc}
\hline
\hline
  & $N_U$   & $N_M$   & $N_L$   & $E_U$   & $E_M$   & $E_L$   \\
\hline
non-Seyfert (68) & 23   & 27   & 18   & 21.9 & 22.4 & 23.7 \\
Seyfert (38)     & 6    & 17   & 15   & 8.2  & 14.1 & 15.7 \\
WTP (15)         & 3    & 10   & 2    & 3.9  & 6.0  & 5.2 \\
\hline
\end{tabular}
\begin{tablenotes}
   \item
\end{tablenotes}
\label{tab:num_uml}
\end{table}

According to the best measurements of CVRs, we counted the numbers of MIR outbursts falling into the three regions ($N_U$, $N_M$ and $N_L$) for non-Seyfert and Seyfert galaxies.
We also calculated the expected value of the numbers in each region ($E_U$, $E_M$ and $E_L$) based on the probability that each source falls in the region: $E_U=\sum_i p_{U,i}$, $E_M=\sum_i p_{M,i}$, and $E_L=\sum_i p_{L,i}$, where $p_{U,i}$, $p_{M,i}$ and $p_{L,i}$ are the probability that the $i$th object falls into the corresponding region.
The results are listed in Table~\ref{tab:num_uml}.
Both calculations yield estimations of the numbers of truth CVR values falling into each region.
Considering the objects in the M region and the objects that were excluded when selecting the rising phase, in the non-Seyfert subsample, $\gtrsim22-23$ are TDEs and $\gtrsim18-24$ are CLAGNs, and in the Seyfert subsample, $\gtrsim6-8$ are ANTs and $\gtrsim15-16$ are CLAGNs.
Therefore, we concluded that the MIRONG sample is a mixture of TDEs/ANTs and CLAGNs, and this is the case for MIR outbursts in both non-Seyfert and Seyfert galaxies.

\subsection{WTP sample} \label{sec:mir_wtp}

The WTP sample contains 18 MIR outbursts, including 12 in their gold subsample and 6 in their silver subsample.
We analyzed their MIR LCs like we did for the MIRONG sample.
We excluded WTP15acbuuv, whose rising phase began at the first data point of NEOWISE-R, and WTP14adbjsh and WTP16aaqrcr, which have only one data point in the rising phase.
Finally, there are 15 objects left for the following analysis.

\begin{figure}
\centering
 \includegraphics[scale=0.66]{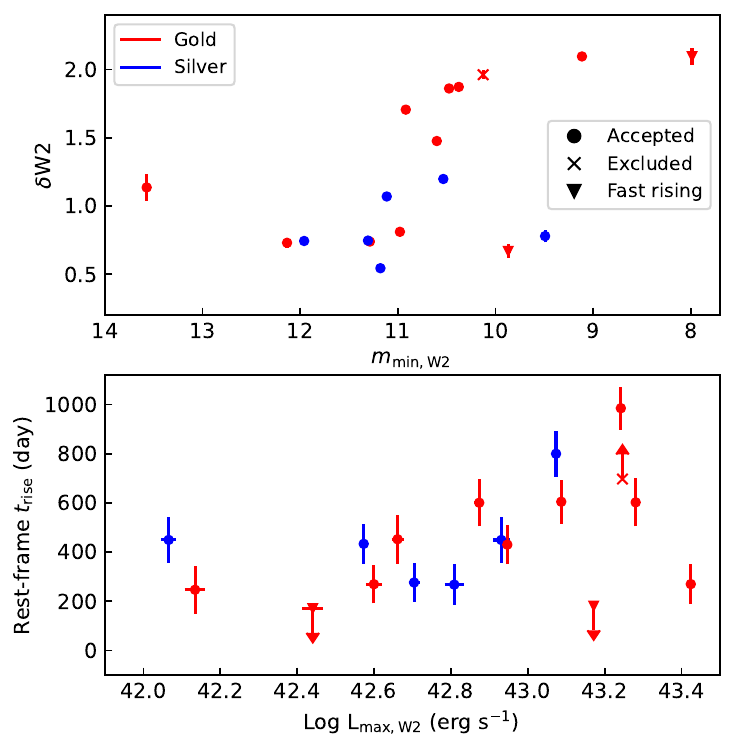}
 \caption{
  MIR properties of the WTP sample.
  The upper panel shows $m_{\rm min,W2}$ and $\delta{\rm W2}$, and the lower panel shows $L_{\rm max,W2}$ and $t_{\rm rise}$.
  The gold and silver subsamples are shown in red and blue symbols.
  The meanings of symbols (dot, cross and inverted triangle) are the same as Figure~\ref{fig:mirproperty_opt}.}
\label{fig:mirproperty_m24}
\end{figure}

The MIR properties of the 15 objects and the three excluded objects are displayed in Figure~\ref{fig:mirproperty_m24}.
The MIR luminosity $L_{\rm max,W2}$ of the sample is in the range of $\sim10^{42}-10^{43.5}$ erg s$^{-1}$, lower than the optically selected sample.
This may be because M24 limited the distance to $<$200 Mpc for the sample.
The rise time $t_{\rm rise}$ is in the range of $<200$ and $\sim1000$ days, shorter than the optically selected sample.
This is because M24 required a rapid rise in the MIR LC.

\begin{figure}
\centering
 \includegraphics[scale=0.66]{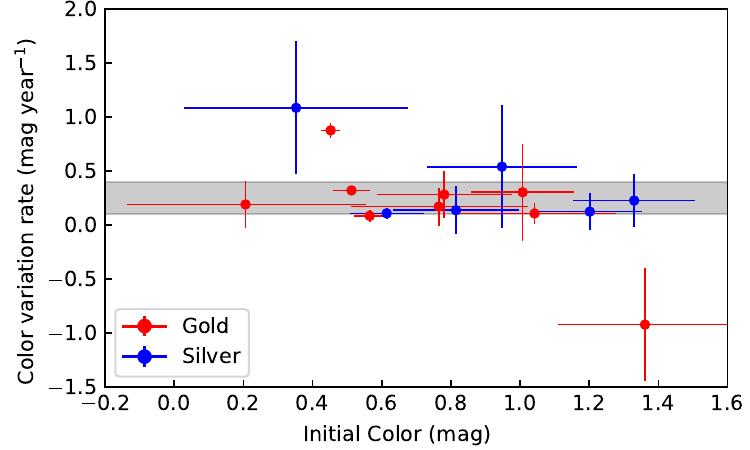}
 \caption{
  The distribution of ICs and CVRs of the WTP sample.
  Red and blue represent the gold and silver subsamples, respectively.}
\label{fig:cvr_ic_m24}
\end{figure}

For the 15 MIR outbursts, we calculated IC, CVR, and the probabilities in the three regions and list the results in Appendix \ref{appendix:param_mir}.
We show the distribution of ICs and CVRs in Figure~\ref{fig:cvr_ic_m24}.
We found no statistical differences in the parameters of gold and silver subsamples divided by M24, probably because there is no intrinsic difference or the sample sizes are too small, so we did not distinguish the two subsamples in the following analysis.
The median CVR of the WTP sample is 0.19 mag yr$^{-1}$, between those of optically selected TDEs (0.44) and CLAGNs (0.09).

We also counted the number of outbursts in the three regions and calculated the expected value of numbers in the same way as we did for the MIRONG sample, and list the results in Table~\ref{tab:num_uml}.
In the WTP sample, $\gtrsim3-4$ are TDEs, and $\gtrsim2-5$ are CLAGNs.
Therefore, we concluded that the WTP sample is also a mixture of TDEs and CLAGNs.

\subsection{Promising TDE candidates} \label{sec:mir_candidate}

In this subsection, we selected promising TDE candidates for use in subsequent studies.

\begin{table*}
\caption{Promising TDE candidates.}
\centering
\begin{tabular}{ccccccccccc}
\hline
\hline
Object          & RedShift & Sample        & CVR   & IC   & $p_U$   & $m_{\rm min,W2}$ & $\delta$W2 & log $L_{\rm max,W2}$    & $t_{\rm rise}$  \\
\hline
WTP17aamoxe     & 0.042    & WTP(Gold)     & $0.88 \pm 0.07 $ & $0.45 \pm 0.03 $ &100.00\% & 10.38 & 1.87 & 43.42 & $270\pm81$ \\
SDSS J0837+4143 & 0.09806  & MIRONG(Comp)  & $1.45 \pm 0.80 $ & $0.54 \pm 0.29 $ & 90.63\% & 12.15 & 0.69 & 43.20 & $236\pm90$ \\
SDSS J1513+3111 & 0.07181  & MIRONG(Comp)  & $0.65 \pm 0.22 $ & $0.45 \pm 0.10 $ & 86.13\% & 12.17 & 1.24 & 43.10 & $260\pm78$ \\
SDSS J1053+5524 & 0.15174  & MIRONG(SF)    & $1.20 \pm 0.74 $ & $0.94 \pm 0.17 $ & 85.49\% & 13.10 & 0.60 & 43.20 & $227\pm86$ \\
SDSS J1628+4810 & 0.12454  & MIRONG(SF)    & $1.14 \pm 0.80 $ & $0.69 \pm 0.27 $ & 82.26\% & 13.61 & 0.63 & 42.82 & $237\pm87$ \\
SDSS J1409+1057 & 0.05972  & MIRONG(LINER) & $0.93 \pm 0.65 $ & $0.35 \pm 0.24 $ & 79.56\% & 11.53 & 0.50 & 42.83 & $251\pm88$ \\
SDSS J1447+4023 & 0.13025  & MIRONG(LINER) & $0.49 \pm 0.14 $ & $0.37 \pm 0.05 $ & 73.95\% & 12.39 & 1.26 & 43.57 & $245\pm76$ \\
SDSS J1524+5314 & 0.08513  & MIRONG(SF)    & $0.54 \pm 0.26 $ & $0.36 \pm 0.09 $ & 70.91\% & 13.40 & 1.08 & 42.73 & $246\pm86$ \\
\hline
SDSS J0842+2357 & 0.06353  & MIRONG(SF)    & $-$              & $0.74 \pm 0.11 $ & -       & 12.80 & 0.81 & 42.60 & $<154$ \\
SDSS J1549+3327 & 0.08565  & MIRONG(SF)    & $-$              & $0.75 \pm 0.03 $ & -       & 12.68 & 2.21 & 43.17 & $<160$ \\
WTP14adbjsh     & 0.0106   & WTP(Gold)     & $-$              & $0.74 \pm 0.03 $ & -       & 7.99  & 2.10 & 43.17 & $<182$ \\
\hline
\end{tabular}
\begin{tablenotes}
   \item
\end{tablenotes}
\label{tab:tde_candidate}
\end{table*}

We selected MIR outbursts in MIRONG and WTP samples that have a probability of falling into U region $p_U$ of $>70\%$, and are hosted in non-Seyfert galaxies.
There are eight outbursts, seven from the MIRONG sample and one from the WTP sample.
We list the MIR information of these TDE candidates in Table~\ref{tab:tde_candidate}.

There should be more TDEs in the MIR samples than these candidates.
For example, if the $p_U$ criterion is applied to the six optical TDEs whose CVRs can be measured, only two (AT 2019dsg and AT 2020nov) can pass.
In addition, the MIR samples are fainter on average, leading to larger measurement errors of CVRs, and hence, it is more difficult to pass the $p_U$ threshold.

According to the analysis of optically selected samples, the MIR emission of TDEs generally has a shorter rise time $t_{\rm rise}$ than CLAGNs.
In particular, the two objects (AT 2017gge and AT 2018dyk) whose CVRs can not be measured due to too short $t_{\rm rise}$ are TDE/ANT instead of CLAGN.
This suggests that objects with short $t_{\rm rise}$ have a high probability of being TDE/ANT.
Thus, we selected three MIR outbursts hosted in non-Seyfert galaxies whose $t_{\rm rise}$ is shorter than the WISE sampling interval.
Their MIR information is also listed in Table~\ref{tab:tde_candidate}.

The MIR luminosities $L_{\rm max,W2}$ of these 11 TDE candidates are in the range of $\sim10^{42.6}-10^{43.6}$ erg s$^{-1}$, similar with optical TDEs with bright MIR emission.
The rise times $t_{\rm rise}$ are generally less than 1 year, shorter than optical TDEs on average.
This might be due to a selection effect, as we found a possible trend that TDEs with shorter rise times have larger CVRs.
For example, the two optical TDEs with the largest CVRs which can pass the $p_U$ criterion, including AT 2019dsg ($t_{\rm rise}=251\pm95$ day, ${\rm CVR}=0.66\pm0.15$) and AT 2020nov ($t_{\rm rise}=262\pm75$ day, ${\rm CVR}=0.74\pm0.23$), also have short rise time.
However, the sample size is not large enough to confirm this trend.

\section{Discussions} \label{sec:discussion}

\subsection{Robustness of our method for distinguishing TDEs/ANTs and CLAGNs} \label{sec:discuss_robustness}

\subsubsection{Possible systematic errors}

We had developed a new method to distinguish TDEs/ANTs from CLAGNs using MIR color variation in the rising phase.
In this subsection we discussed the uncertainties of the method.

The systematic errors of CVR measurement may come from the following aspects.
First, the K-correction of the MIR color has a systematic error of $\sigma_{\rm sys}=0.04 z$ mag due to uncertain SEDs, as estimated in Appendix~\ref{appendix:kcorr}.
Most of the objects in the samples in this work are at redshifts below 0.2, leading to systematic errors of $\lesssim0.08$ mag.
Secondly, our simulations indicate that the difference in sampling time may lead to a scatter in measured CVR of 0.02--0.15 mag yr$^{-1}$ (standard deviation) for the same MIR outburst, as shown in Figure \ref{fig:simu_uvlc_shape}, \ref{fig:simu_lagn}, \ref{fig:simu_lmax_tau} and \ref{fig:simu_dust_shape}.
Finally, for a small fraction of MIR outbursts with poor data quality, some data points in the rising phase may not be included when calculating CVR because they do not meet the 3$\sigma$ criterion in the W1 or W2 band.

The above systematic errors caused the CVR errors we gave to underestimate the actual uncertainty.
Fortunately, the systematic errors should not significantly affect our main results.
For most objects, the statistical errors are larger than the systematic errors and dominate the total errors.
Moreover, as we demonstrated in section \ref{sec:mircolor}, measurement errors make the CVR distribution of a sample more diffused and blur the CVR distinctions between samples.
Therefore, underestimating measurement error does not invalidate the conclusion that there is a CVR difference between TDEs/ANTs and CLAGNs based on optical samples, nor does it affect the reliability of CVR thresholds to divide regions.

Our conclusions and methods rely on the optical sample containing 13 TDE+ANTs and 19 CLAGNs.
We cannot rule out the existence of ``anomalies'' whose CVRs deviate significantly from the usual parameter range of the sample.
Considering these possible anomalies, high $p_U$ ($p_L$) alone does not necessarily mean that a MIR outburst is TDE/ANT (CLAGN).
For example, we considered the probability $p({\rm TA}|U)$ that an object with CVR falling in the U region is TDE/ANT according to Bayes' equation:
\begin{equation}
p({\rm TA}|U) = \frac{ f_{\rm TA} p(U|{\rm TA}) }{ f_{\rm TA} p(U|{\rm TA}) + f_{\rm CLAGN} p(U|{\rm CLAGN}) },
\end{equation}
where $f_{\rm TA}$ and $f_{\rm CLAGN}=1-f_{\rm TA}$ are the fractions of TDE/ANT and CLAGN in the MIR outburst sample, respectively, and $p(U|{\rm TA})$ or $p(U|{\rm CLAGN})$ is the probability that the CVR of a TDE/ANT or CLAGN falls in the U region.
We adopted a $p(U|{\rm TA})$ value of 7/13 based on the observed CVRs of optical TDE+ANTs, and constrained $p(U|{\rm CLAGN})$, the fraction of anomalies in CLAGNs, to be less than 1/19 based on the number of optical CLAGNs.
Provided that the amounts of TDE+ANTs and CLAGNs in MIR outbursts are comparable ($f_{\rm TA}$ in the range of 0.3--0.8, see section \ref{sec:discuss_tde_fraction}), $p(TA|U)$ is higher than 82\%--98\%.
Based on a similar analysis, the probability that an object with CVR falling in the L region is CLAGN is higher than 77\%--95\%.
Therefore, considering possible anomalies leads to a slight decrease in the probability that an outburst with high $p_U$ ($p_L$) is TDE/ANT (CLAGN).

\subsubsection{Comparison with diagnoses from other bands}

We tested our method for distinguishing TDEs, ANTs, and CLAGNs by comparing our classifications for MIR outbursts with diagnoses using data from other bands in the literature.

\begin{table*}
\caption{Comparison of our and literature diagnoses of MIR outbursts}
\begin{tabular}{ccccc}
\hline
\hline
Object          & Our & Notes  & Literature  & Literature notes  \\
\hline
\multicolumn{5}{c}{\citet{Wang2022_MIRONG} non-Seyfert} \\
\hline
J1513+3111   & TDE    & ${\rm CVR}=0.65\pm0.22$, $p_U=86.1\%$ & TDE       & declining, iron CL, He II, N III \\
J1549+3327   & TDE    & $t_{\rm rise}<160$ day                & TDE       & declining, iron CL, He II \\
J1043+2716   & CLAGN  & ${\rm CVR}=0.04\pm0.09$, $p_L=74.9\%$ & TDE       & declining, iron CL \\
J1442+5558   & CLAGN  & ${\rm CVR}=0.00\pm0.12$, $p_L=80.6\%$ & turn-on   & maintain, iron CL, He II, N III \\
\hline
\multicolumn{5}{c}{\citet{Wang2022_MIRONG} Seyfert} \\
\hline
J1332+2036   & ANT    & ${\rm CVR}=1.00\pm0.12$, $p_U=100\%$  & AGN flare & declining \\
J1402+3922   & ANT    & ${\rm CVR}=1.08\pm0.18$, $p_U=100\%$  & TDE       & restored, iron CL, He II \\
\hline
\multicolumn{5}{c}{\citet{Masterson2024} with X-ray} \\
\hline
%WTP 15abymdq & unsure & ${\rm CVR}=0.19\pm0.22$               & unsure    & hard X-ray with $\Gamma<2.0$ \\
WTP 17aamoxe & TDE    & ${\rm CVR}=0.88\pm0.07$, $p_U=100\%$  & TDE       & soft X-ray with $\Gamma=3.8^{+1.4}_{-1.9}$ \\
WTP 17aamzew & unsure & ${\rm CVR}=0.28\pm0.22$               & TDE       & soft X-ray with $\Gamma=4.1^{+2.3}_{-1.8}$ \\
\hline
\end{tabular}
\begin{tablenotes}
   \item
\end{tablenotes}
\label{tab:compare_other_band}
\end{table*}

\citet[][hereafter Wang22]{Wang2022_MIRONG} carried out follow-up optical spectral monitoring of 22 MIR outbursts in the MIRONG sample and diagnosed their nature.
We ignored two optical TDEs and one optical CLAGN, and of the remaining 19 MIR outbursts, 10 are in non-Seyfert galaxies, and nine are in Seyfert galaxies.
We selected four MIR outbursts with a short rise time or $p_U>70\%$ as promising TDEs/ANTs (two TDEs in non-Seyfert galaxies and two ANTs in Seyfert galaxies), and two with $p_L>70\%$ as promising CLAGNs.
They are listed in Table~\ref{tab:compare_other_band}.

We compared our diagnoses with those by Wang22, which are also listed in Table~\ref{tab:compare_other_band}.
We found that except for some differences in terms (for example, we referred to TDE-like flares in AGNs as ANTs without distinguishing their natures), only one out of the six was classified differently: J1043+2716 was classified as CLAGN with $\gtrsim75\%$ confidence by us while classified as a TDE by Wang22 due to its declining broad emission lines and the presence of iron coronal lines.
The MIR outburst J1043+2716 was first detected in May 2015, and the optical outburst must have begun earlier.
Wang22's classification relied on four spectra taken between April 2017 and March 2021, $\sim2-5$ years after the outburst began.
Without early-time spectra, the possibility of CLAGN cannot be ruled out by declining broad emission lines.
In addition, strong iron coronal emission lines are also detected in AGNs \citep{Clark2024}, so the possibility of CLAGN cannot be fully ruled out by coronal lines.
Besides, the probabilities of our promising TDE candidates are not high enough, and there may be individual sources that were misclassified.
Therefore, the conflict between our and Wang22's classification of J1043+2716 could be reconciled.

X-ray spectra are a reliable basis for distinguishing between TDEs and CLAGNs: many TDEs have soft spectra with steep spectral indices $\Gamma>2$ \citep[e.g.,][]{Saxton2020}.
In contrast, AGNs have relatively hard spectra with $\Gamma \lesssim 2$ \citep[e.g.,][]{Piconcelli2005}.
In the WTP sample, three MIR outbursts were detected in the X-ray band by eROSITA after they occurred.
Based on soft and steep X-ray spectra, \citet{Masterson2024} identified two likely TDEs, whose information is listed in Table~\ref{tab:compare_other_band}.
Of these two X-ray-verified TDEs, WTP 17aamoxe is also identified as likely TDE by us because it falls in the U region with 100\% probability, and WTP 17aamzew has a CVR of $0.28\pm0.22$, also consistent with the TDE interpretation.
Therefore, our classifications for the WTP sample agree with the X-ray diagnoses.

\subsection{Fractions of TDEs, ANTs and CLAGNs in MIR selected samples} \label{sec:discuss_tde_fraction}

In sections \ref{sec:mir_mirong} and \ref{sec:mir_wtp}, we estimated the lower limits of the numbers of TDEs, ANTs and CLAGN using the numbers in the U and L regions for MIRONG and WTP samples.
However, the numbers do not include objects in the M region, where individual objects could not be reliably classified.
Nevertheless, we could estimate the numbers of TDEs, ANTs and CLAGNs in the M region by using the probabilities of these types of outbursts falling in the M region.

For the eight optical TDEs, six have CVRs in the U region, and two have CVRs in the M region.
Assuming that the fraction of TDEs in the U region $f_{\rm U,TDE}$ has a uniform prior distribution between 0 and 1, and the observed number obeys the binomial distribution, we calculated a posterior distribution using Bayes' method.
Using the posterior distribution, we obtained $f_{\rm U,TDE}=0.71^{+0.13}_{-0.16}$ and $f_{\rm M,TDE}=1-f_{\rm U,TDE}=0.29^{+0.16}_{-0.13}$.
In a similar way, we obtained $f_{\rm U,ANT}=0.32^{+0.17}_{-0.14}$, $f_{\rm M,ANT}=0.68^{+0.14}_{-0.17}$, $f_{\rm L,CLAGN}=0.57\pm0.11$, and $f_{\rm M,CLAGN}=0.43\pm0.11$.

For the MIRONG non-Seyfert sample with $N_U=23$, $N_M=27$ and $N_L=18$, we assumed that the number of TDEs in the U region is $N_U$ and the number of CLAGNs in the L region is $N_L$.
Then, the numbers of TDEs and CLAGNs in the M region satisfy $N_{\rm M,TDE} = N_U \times \frac{ f_{\rm M,TDE} }{ f_{\rm U,TDE} }$ and $N_{\rm M,CLAGN} = N_L \times \frac{ f_{\rm M, CLAGN} }{ f_{\rm L,CLAGN} }$.
Thus, their probability density distributions are:
\begin{equation}
\begin{aligned}
\frac{{\rm d}p(N_{\rm M,TDE})}{{\rm d}N_{\rm M,TDE}} =& \frac{{\rm d}p(f_{\rm U,TDE})}{{\rm d}f_{\rm U,TDE}}  \frac{f_{\rm U,TDE}^2}{N_U} \\
\frac{{\rm d}p(N_{\rm M,CLAGN})}{{\rm d}N_{\rm M,CLAGN}} =& \frac{{\rm d}p(f_{\rm L,CLAGN})}{{\rm d}f_{\rm L,CLAGN}}  \frac{f_{\rm L,CLAGN}^2}{N_L}
\end{aligned}
\end{equation}
With the constraint equation $N_{\rm M,TDE}+N_{\rm M,CLAGN}=N_M$, we calculated the posterior probability distributions of $N_{\rm M,TDE}$ and $N_{\rm M,CLAGN}$ using Bayes' equation.
The results show that $N_{\rm M,TDE}=11.3\pm5.1$ and $N_{\rm M,CLAGN}=15.7\pm5.1$.
In a similar way, we obtained $N_{\rm M,ANT}=7.7\pm2.7$ and $N_{\rm M,CLAGN}=9.3\pm2.7$ for the MIRONG Seyfert sample, and $N_{\rm M,TDE}=8.0^{+0.9}_{-1.8}$ and $N_{\rm M,CLAGN}=2.0^{+1.8}_{-0.9}$ for the WTP sample.

We then considered the MIR outbursts that were excluded during the selection of the rising phase.
There are two cases: one is that the $t_{\rm rise}$ is too short to calculate CVR, and the other is that the outburst starts at the first data point of NEOWISE-R.
We assumed that outbursts in the former case are all TDEs/ANTs.
The outbursts in the latter case could not be classified using color variation, so they are either TDE/ANT or CLAGN.
With these outbursts included, we estimated the fractions of TDEs/ANTs and CLAGNs in the MIR outburst samples, and the results are listed in Table~\ref{tab:frac_tde_clagn}.

\begin{table}
\centering
\caption{Fractions of TDEs/ANTs in the MIR outburst samples.}
\begin{tabular}{cccc}
\hline
\hline
  & $f_{\rm TDE}$   & $f_{\rm ANT}$   & $f_{\rm CLAGN}$  \\
\hline
non-Seyfert (77) & 45\%--63\% & -          & 37\%--55\% \\
Seyfert (50)     & -          & 30\%--57\% & 43\%--70\% \\
WTP (18)         & 62\%--82\% & -          & 18\%--38\% \\
\hline
\end{tabular}
\begin{tablenotes}
   \item
\end{tablenotes}
\label{tab:frac_tde_clagn}
\end{table}

The results confirmed our conclusion in section \ref{sec:mir_sample} that all MIR outburst samples are mixtures of TDEs/ANTs and CLAGNs.
Overall, MIR outbursts consist of comparable amounts of TDEs/ANTs and CLAGNs: more than half in non-Seyfert galaxies are TDEs, and more than half in Seyfert galaxies are CLAGNs.

Can CLAGNs occur in non-Seyfert galaxies?
Non-Seyfert galaxies were selected using narrow-line-ratio and MIR color diagnoses \citep{Jiang2021,Masterson2024}.
These diagnoses cannot fully ruled out weak AGNs whose signals are overwhelmed by star formation.
Thus, for the outbursts in the MIRONG and WTP non-Seyfert samples that were diagnosed as CLAGNs by MIR color, their host galaxies may have weak AGNs, even though there was no clear AGN signal before the outbursts.

The fraction of TDEs in the WTP sample (62\%--82\%) is higher than that in the MIRONG non-Seyfert sample (45\%--63\%).
This is possibly because \citet{Masterson2024} required an additional criterion that the MIR LC must have a fast rise and a slow, monotonic decay.
However, CLAGNs could not be completely excluded after adding this criterion, as the fraction of CLAGNs in the WTP sample is still $\gtrsim20\%$, possibly because some CLAGNs also meet this criterion.
In addition, it is likely that some TDEs, for example, AT 2019qiz (Figure~\ref{fig:example_ATs}) whose MIR emission rises nonmonotonically in $>925$ days, fail to pass this criterion.
Therefore, this standard may improve the purity of TDEs while losing completeness.

\subsection{Implications and future works} \label{sec:discuss_future}

\citet{Jiang2021} and \citet{Masterson2024} obtained incident rates of MIR outbursts with their samples of $5.4\times10^{-5}$ and $2.0\pm0.3\times10^{-5}$ galaxy$^{-1}$ yr$^{-1}$, respectively.
We estimated the incident of IR TDEs using their results and our estimations of the TDE fractions.
For the MIRONG sample, with the 77 MIR outbursts in non-Seyfert galaxies that are not supernovae and are radio-quiet, we could estimate a rate of $3.4\times10^{-5}$ galaxy$^{-1}$ yr$^{-1}$ considering that non-Seyfert galaxies make up $\sim90\%$ of SDSS galaxies \citep{Kewley2006}.
Assuming that 45\%--63\% of them are TDEs, the incident rate of IR TDEs is $1.5-2.1\times10^{-5}$ galaxy$^{-1}$ yr$^{-1}$.
For the WTP sample, \citet{Masterson2024} estimated the rate using 12 MIR outbursts in their gold sample.
According to our estimation, 62\%--82\% of their total 18 outbursts are TDEs, and hence the rate should be modified to $1.6-2.8\times10^{-5}$ galaxy$^{-1}$ yr$^{-1}$, which is consistent with the rate \citet{Masterson2024} derived using only their gold sample.
Therefore, the results from the MIRONG and WTP samples are consistent, and the final rate is $1.5-2.8\times10^{-5}$ galaxy$^{-1}$ yr$^{-1}$, comparable with the optical TDE rate of $3.2^{+0.8}_{-0.6}\times10^{-5}$ galaxy$^{-1}$ yr$^{-1}$ \citep{Yao2023} within the margin of error.
%However, the WTP sample may missed IR TDEs with long rise time like AT 2019qiz.
%Assuming a fraction of $\sim1/7$ missed, the corrected rate is $1.5-1.9\times10^{-5}$ galaxy$^{-1}$ yr$^{-1}$.

Part of IR TDEs are bright in both optical and IR bands and are also included in optical TDEs.
\citet{Jiang2021} and \citet{Masterson2024} estimated optical detections for MIR outbursts of 11\% and 22\%, respectively.
In addition, of the 33 optical TDEs in \citet{Yao2023}, only two (AT 2019dsg and AT 2019qiz) are MIR bright.
Thus, there is not much overlap between optical TDEs and IR TDEs.
Combining results from optical, X-ray and IR surveys, we estimated a total TDE rate of $\sim5-8\times10^{-5}$ galaxy$^{-1}$ yr$^{-1}$.

So far, our method has not been able to classify most of the MIR outbursts clearly, mainly due to the large measurement errors of MIR colors and the semi-annual sampling interval of the WISE project, which is not frequent enough.
Nevertheless, some MIR outbursts can already be classified based on the available data.
For example, SDSS J154843.06+220812.6, whose radio counterpart VT J154843.06+220812.6 has been studied in detail by \citet{Somalwar2022}, is a likely CLAGN because of its CVR of $-0.08\pm0.03$, corresponding a $p_L$ of 100\%.

Although the WISE telescope was retired in July 2024, the Near-Earth Object Surveyor \citep{Mainzer2023} will launch in 2027.
Furthermore, there will be some wide-field NIR survey projects in the future, such as the wide-field infrared transient explorer \citep[WINTER,][]{Lourie2020_winter}, PRime-focus Infrared Microlensing Experiment \citep[PRIME,][]{Kondo2023_prime}, and the Nancy Grace Roman Space Telescope \citep{Akeson2019_wfirst}.
These projects will discover a large amount of IR outbursts in galaxy centers, and obtain their high-quality, high-cadence NIR LCs.
Our simulations predict that the NIR color $H-K$ of TDEs also turns red faster in the rising phase than CLAGNs, with a pattern similar to MIR color (Appendix \ref{appendix:simu_nir}).
If true, IR outbursts can also be classified based on the NIR CVR.
Due to the high quality and cadence of NIR data, we expected to obtain more IR outbursts with precise classification by the NIR CVR.

We have found that the rise times and variability amplitudes of TDEs and CLAGNs are also different on average.
Although it is difficult to distinguish TDEs and CLAGNs using each parameter itself, by combining all the parameters with the help of machine learning, the accuracy of classification may be improved.

The classification based on CVR has a unique advantage in that a preliminary diagnosis can be obtained using only the rising phase of the LC.
This allows one to select likely TDEs or exclude likely CLAGNs in the early phases of IR outbursts and is further convenient for timely multi-band follow-up observations.

\section{Summary and conclusions} \label{sec:summary}

In this work, we selected MIR-bright optical outbursts, including 9 TDEs, 8 ANTs and 19 CLAGNs.
Using the WISE MIR LCs sampled every half a year since 2014 in W1 and W2 bands, we calculated the K-corrected W1-W2 color in the rising phase and investigated its time variation.

We found that the MIR color of TDEs and ANTs generally turns red quickly during the rising phase, while CLAGN generally does not.
Quantitatively, the median CVR of TDEs and ANTs is $\sim0.4$ mag yr$^{-1}$, higher than that of CLAGNs of 0.09 mag yr$^{-1}$.
The CVR distributions of TDEs and ANTs differ from that of CLAGNs by significances of $2.9\times10^{-5}$ and $3.1\times10^{-4}$, respectively, estimated using K-S tests.
In addition, the initial MIR color of TDEs is generally redder than that of ANTs, as the median ICs of TDEs and ANTs are 0.52 and 0.36 mag, respectively, which differ by a significance of 0.020.

Using simulations involving the dust radiative transfer model, we explored the origins of the two phenomena.
One phenomenon, the MIR color of TDEs and ANTs turning red quickly, is possibly because of their UV LCs characterized by ``rapid rise, short peak and long tail'', and/or because of no or weak contribution from underlying AGN.
The other one, ANTs showing bluer initial colors than TDEs, is possibly because of higher fraction of graphite and larger grain size of the echoing dust, and/or because of the difference in dust geometry: dust around ANTs has a torus shape with face-on inclination, while that around TDEs is approximately spherically symmetric.

We proposed a method to classify MIR outbursts based on the CVR difference between TDEs/ANTs and CLAGNs.
We divided three regions: a U region with CVR greater than 0.4 mag yr$^{-1}$ representing a high probability of being TDE/ANT, an L region with CVR less than 0.1 mag yr$^{-1}$ representing a high probability of being CLAGN, and an M region between them, where the outbursts are difficult to classify.
In addition, MIR outbursts with a rise time shorter than half a year have a high probability of being TDEs/ANTs, because the two optical outbursts that meet this condition are TDE/ANT.

Using this method, we classified 127 MIR outbursts in the MIRONG sample of \citet{Jiang2021} and 18 in the WTP sample of \citet{Masterson2024}.
For the MIRONG sample, we divided it into a non-Seyfert subsample and a Seyfert subsample, according to the types of host galaxies, containing 77 and 50 MIR outbursts, respectively.
We found that in all three samples, both likely TDEs/ANTs and likely CLAGNs exist, and hence, they are all mixtures of TDEs/ANTs and CLAGNs.
We selected 11 promising TDE candidates based on a high probability of CVR falling in the U region or a short rise time.
We examined our classifications with those in the literature based on follow-up optical spectra and X-ray observations and found no apparent contradiction.

We inferred that 45\%--63\% of MIR outbursts are TDEs in the MIRONG non-Seyfert subsample, 30\%-57\% in the MIRONG Seyfert subsample are ANTs, 62\%--82\% in the WTP sample are TDEs, and the remainders are CLAGNs.
Thus, MIR outbursts contain comparable amounts of TDEs/ANTs and CLAGNs overall, and those in non-Seyfert galaxies are mostly TDEs.
Using the TDE fractions above, we estimated the incident rate of IR TDEs in non-Seyfert galaxies to be $1.5-2.8\times10^{-5}$ galaxy$^{-1}$ yr$^{-1}$ with the MIRONG and WTP samples, comparable with that of optical TDEs within the margin of error.

Although this work only studied the variation of MIR color, it is hopeful that the variation of NIR color can also be used to classify IR outbursts.
We expect that more IR outbursts will be discovered and classified with future high-quality, high-cadence IR surveys.
This will eventually shed new light on the statistical study of obscuring TDEs.

\begin{acknowledgements}
This work is supported by the National Natural Science Foundation of China (NFSC, 12103002) and the University Annual Scientific Research Plan of Anhui Province (2022AH010013, 2023AH030052).
This publication makes use of data products from NEOWISE, which is a project of the Jet Propulsion Laboratory/California Institute of Technology, funded by the Planetary Science Division of the National Aeronautics and Space Administration.
This research has made use of the NASA/IPAC Infrared Science Archive, which is operated by the California Institute of Technology, under contract with the National Aeronautics and Space Administration,
the archival data are \href{https://doi.org/10.26131/irsa535}{DOI: 10.26131/IRSA535} \citep{https://doi.org/10.26131/irsa535} and \href{https://doi.org/10.26131/irsa144}{DOI: 10.26131/IRSA144} \citep{https://doi.org/10.26131/irsa144}.

\textit{Software}: This work made use of the following \texttt{python} packages:
\texttt{Numpy} \citep{numpy}, \texttt{Matplotlib} \citep{matplotlib}, \texttt{Scipy} \citep{scipy} \texttt{Astropy} \citep{astropy}, and \texttt{emcee} \citep{emcee}.

\end{acknowledgements}

\bibliography{tde_clagn_mircolor}

\begin{appendix}

\section{Display of special objects} \label{appendix:special}

\setcounter{figure}{0}
\renewcommand{\thefigure}{A\arabic{figure}}
\renewcommand*{\theHfigure}{\thefigure}

\begin{figure}
\centering
 \includegraphics[scale=0.88]{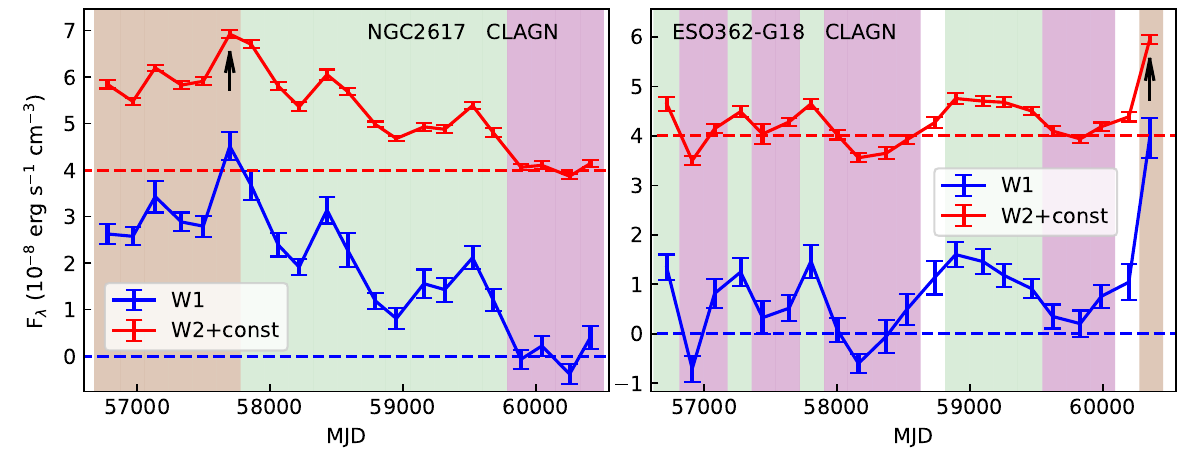}
 \caption{
  The MIR LCs, the selections of quiescent, outburst states and rising phase of NGC 2617 and ESO362-G18.
  The meaning of labels and colors are the same as Figure~\ref{fig:example_quie_outb}(b).}
  \label{fig:example_excluded}
\end{figure}

%%% Optical CLAGNs excluded when selecting the rising phase
Here, we show the information of the eight CLAGNs excluded in section~\ref{sec:opt_rising} and demonstrate why we excluded them.
For seven CLAGNs, including 3C 390.3, HE1136-2304, Mrk 6, Mrk 926, NGC 2617, NGC 4395, and SDSS J080020.98+263648.8, the outburst started in the first data point of NEOWISE-R, which are typically between December 2013 and June 2014.
We show an example NGC 2617 (Figure~\ref{fig:example_excluded}(a)), for which our algorithm identified an outburst from 2014 to 2022.
However, the outburst started earlier than April 2013 \citep{Shappee2014}.
Therefore, it is likely that NEOWISE-R observations missed the beginning of the outburst for the seven CLAGNs.
For one CLAGN, ESO 362-G18 (Figure~\ref{fig:example_excluded}(b)), the outburst started too late, and only one data point can be used in the rising phase.
The color variation rate in the rising phase can not be calculated, so we removed it.

%%% AT 2017gge and AT 2018bcb
%%% AT 2019qiz

\begin{figure}
\centering
 \includegraphics[scale=0.88]{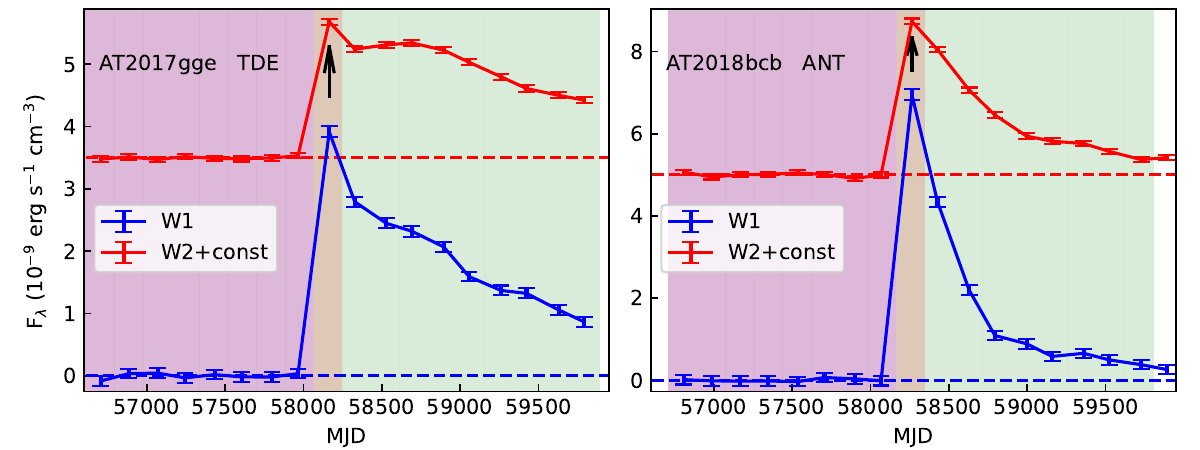}
 \includegraphics[scale=0.88]{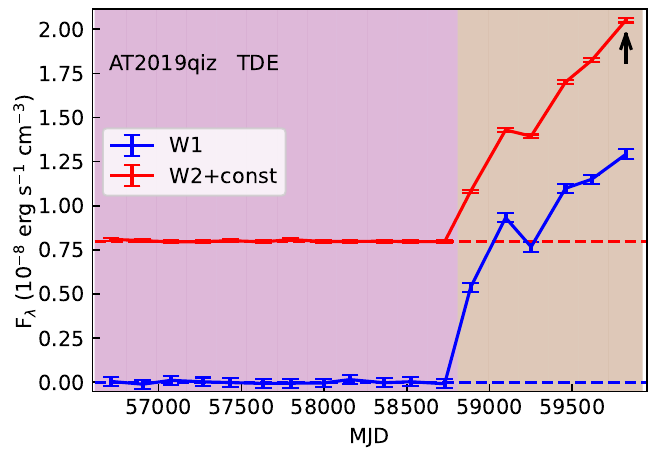}
 \caption{
  The MIR LCs, the selections of quiescent, outburst states and rising phase of AT 2017gge, AT 2018bcb and AT 2019qiz.
  The meaning of labels and colors are the same as Figure~\ref{fig:example_quie_outb}(b).}
  \label{fig:example_ATs}
\end{figure}

In addition, we display the MIR LCs of TDE AT 2017gge and ANT AT 2018bcb with short rise time, and that of TDE AT 2019qiz with long rise time.

\section{K-correction for the MIR color} \label{appendix:kcorr}

\setcounter{figure}{0}
\renewcommand{\thefigure}{B\arabic{figure}}
\renewcommand*{\theHfigure}{\thefigure}

\begin{figure}
\centering
 \includegraphics[scale=0.7]{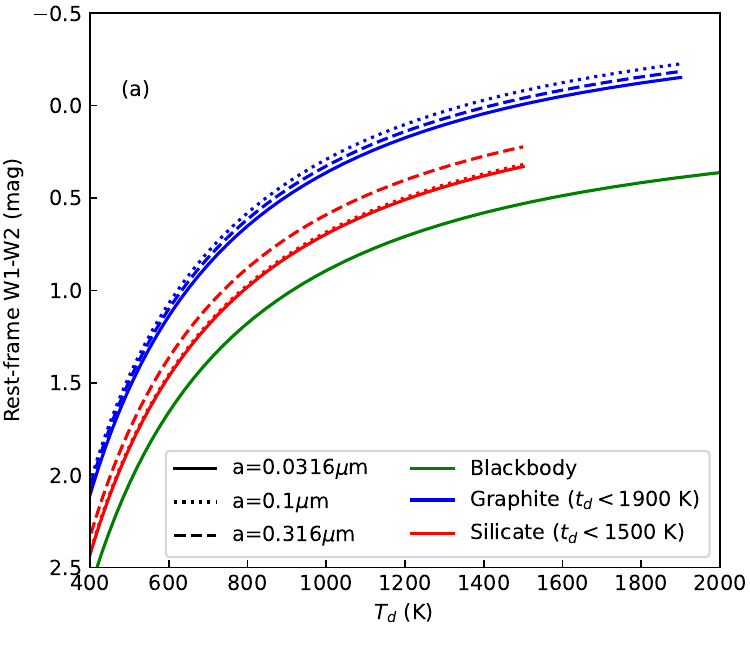}
 \includegraphics[scale=0.7]{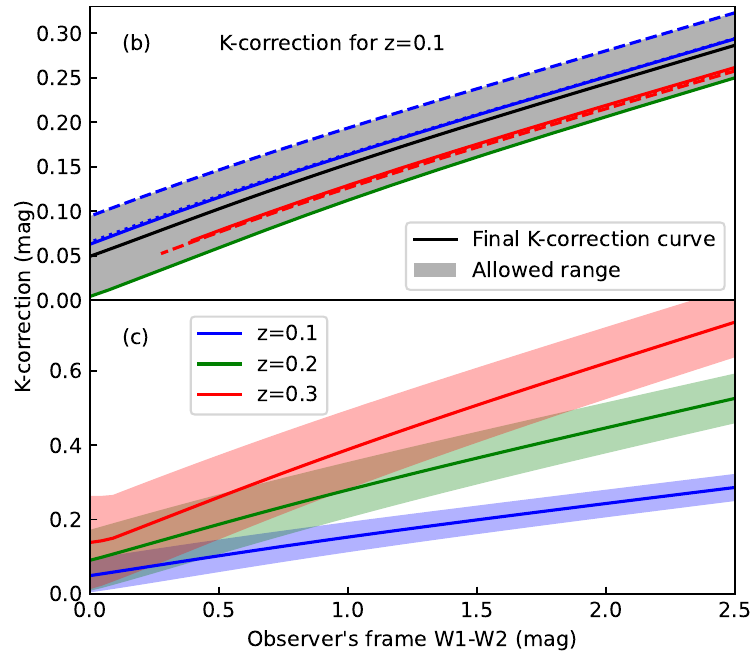}
 \caption{
  \textbf{(a)}: Rest-frame $W1-W2$ color as functions of temperature for different types of SEDs.
  Blue, red and green lines represent results of optically thin graphite and silicate dust, and blackbody.
  For graphite and silicate dust, results assuming grain radii of 0.0316, 0.1 and 0.316 $\mu$m are shown using solid, dotted and dashed lines.
  \textbf{(b)}: The K-correction for $z=0.1$ as functions of color in the observer's frame for different types of SEDs.
  The meanings of colors and styles of the lines are the same as panel (a).
  The grey shades show the allowed range of K-correction, and the black solid line shows the final K-correction curve.
  \textbf{(c)}: The final K-correction curves and allowed range for $z=0.1$, 0.2 and 0.3.
  }
  \label{fig:color_kcorr}
\end{figure}

The color $C=W1-W2$ of a target in the observer's frame is determined by SED and redshift.
We considered three types of SEDs: blackbody, optically thin graphite, and silicate.
The blackbody SED is expressed as:
\begin{equation}
L_\lambda(\lambda) = \pi B_\lambda(\lambda,T_{\rm BB}) 4\pi r_{\rm BB}^2,
\end{equation}
where $B_\lambda$ is the Planck function, $T_{\rm BB}$ and $r_{\rm BB}$ are the blackbody temperature and blackbody radius, respectively.
The SED of optically thin graphite or silicate can be expressed as:
\begin{equation}
L_\lambda(\lambda) = 4\pi B_\lambda(\lambda,T_d) M_d \kappa_\lambda(a),
\end{equation}
where $T_d$ and $M_d$ are the temperature and mass of the dust, and $\kappa_\lambda(a)$ is the dust mass absorption coefficient for grain size $a$.
We adopted the absorption coefficients from Laor \& Draine (1993) and converted them to $\kappa_\lambda$ following Fox et al. (2010).

Using the SEDs, we predicted the color $C$ for a given redshift $z$ as:
\begin{equation}
C(z) = - 2.5\ {\rm Log} \left[ \frac{ \int^\infty_0 f_\lambda(\frac{\lambda}{1+z}) S_{\rm W1}(\lambda) {\rm d}\lambda }{ \int^\infty_0 f_\lambda(\frac{\lambda}{1+z}) S_{\rm W2}(\lambda) {\rm d}\lambda } \  \frac{f_{0,W2}}{f_{0,W1}} \right]   ,
\end{equation}
where $S_{\rm W1}(\lambda)$ and $S_{\rm W2}(\lambda)$ are the normalized transmission curves of the WISE W1 and W2 filters\footnote{https://wise2.ipac.caltech.edu/docs/release/prelim/expsup/sec4\_3g.html}, and $f_{0,W1}$ and $f_{0,W2}$ are the zero points of the two filters.

In Figure~\ref{fig:color_kcorr}(a), we show the relationships between rest-frame colors and temperature for different SEDs.
The results for grain sizes $a=0.0316$, 0.01 and 0.316 $\mu$m are displayed for optically thin graphite and silicate.
At the same temperature, graphite dust is the bluest, followed by silicate dust, and blackbody is the reddest.
The color of actual dust may be more complex.
If the dust is a mixture of graphite and silicate, the color would be between pure graphite and silicate.
In addition, if the MIR optical depth of the dust is on the order of magnitude of 1, the color would be between those of optically thin dust and blackbody.
Therefore, despite the complexity of the SED of actual dust, its color can still be constrained by the above simple SED models.

For different redshifts, temperatures, and types of SEDs, we calculated K-correction $K(z)=C(z)-C(0)$, where $C(z)$ and $C(0)$ are colors in the observer's frame and the rest frame, respectively.
We show the $K(z)$ as a function of $C(z)$ for $z=0.1$ in Figure~\ref{fig:color_kcorr}(b).
We found that the functions assuming different types of SEDs are approximately a series of parallel lines.
For $z<0.35$, the top is always graphite with $a=0.316$ $\mu$m, and the bottom is always the blackbody.
Therefore, we adopted the average of the $K(z)$ given by these two types of SEDs as the final K-correction curve (the black line), and estimated the systematic error by half of their difference.
The final K-corrections and the systematic errors both increase as $z$ increases, as shown in Figure~\ref{fig:color_kcorr}(c).
The systematic error caused by uncertain types of SEDs can be roughly estimated as $\sigma_{\rm sys}(K)\approx0.04 z$.

\section{Analysis of the 2020 outburst of Mrk 1018} \label{appendix:mrk1018}

\setcounter{figure}{0}
\renewcommand{\thefigure}{C\arabic{figure}}
\renewcommand*{\theHfigure}{\thefigure}

\begin{figure}
\centering
 \includegraphics[scale=0.85]{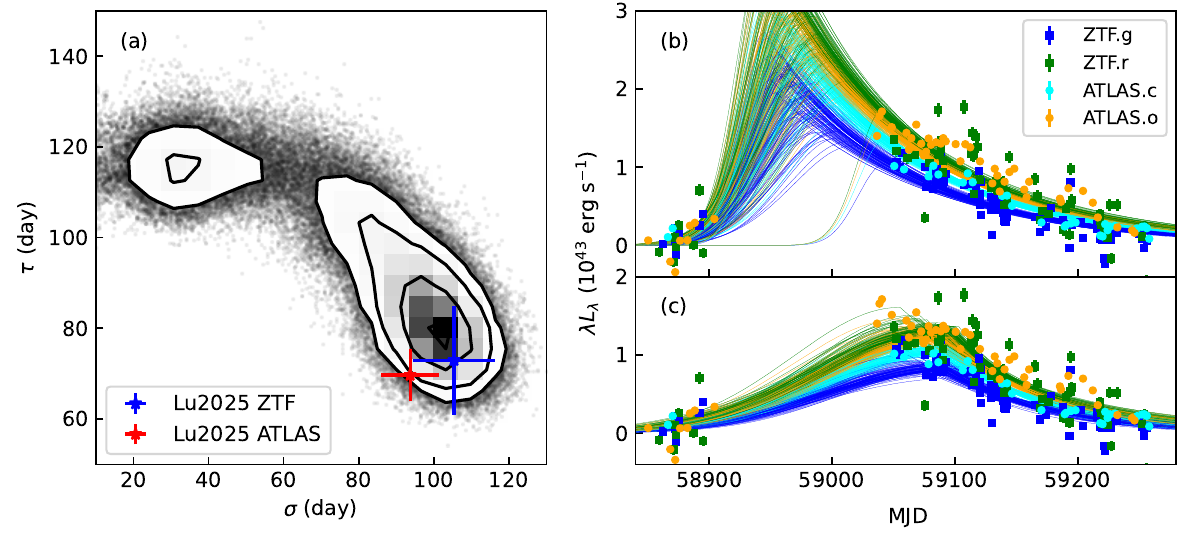}
 \caption{
  \textbf{(a)}: Posterior probability density distributions of $\sigma$ and $\tau$ of the GP model.
  We show the parameters derived from ZTF (blue) and ALTAS (red) data separately by \citet{Lu2025_Mrk1018} with stars for comparison.
  These parameters are consistent with our slowly rising models with higher probability, confirming that the GE models of \citet{Lu2025_Mrk1018} are best-fitting.
  However, they ignored fast-rising models, which are also acceptable.
  \textbf{(b)}: Fast rising models ($\sigma<60$ days) in the ZTF/g (blue), ZTF/r (green), ATLAS/c (cyan) and ATLAS/o (orange) bands and the observed data.
  We show 100 randomly selected MCMC realizations.
  \textbf{(c)}: Same as (b) but show slowly rising models ($\sigma>60$ days).
  }
\label{fig:mrk1018_optlc}
\end{figure}

Mrk1018 is a famous CLAGN.
It changed from type 1.9 to type 1 between 1979 and 1984 \citep{Cohen1986_Mrk1018}, and back to type 1.9 between 2010 and 2015 \citep{Husemann2016_Mrk1018}.
In 2020, it changed to type 1 again, along with a giant outburst lasting for about one year, and then returned to type 1.9 after the outburst ended.

%\citep{Brogan2023_Mrk1018,Lu2025_Mrk1018}

In this work, we are only concerned with the shape of the UV LC of the 2020 outburst.
Does it rise quickly and fall slowly like ANT, or does it have a similar rise time and fall time like most CLAGNs?
This is hard to answer because Mrk 1018 has no observations from March to June 2020, which is the expected rise phase.
\citet{Lu2025_Mrk1018} found that the optical LCs can be fitted equally well with two models: one is a Gaussian model, and the other is a GE model including a Gaussian rise and an exponential fall.
In the GE model, the rise and fall times can be represented by $\sigma$ and $\tau$, respectively.
Thus, comparing $\sigma$ and $\tau$ tells one what type of LC it is.
Using ZTF and ATLAS LCs, \citet{Lu2025_Mrk1018} yielded best-fitting parameters of $\sigma=\sim90-110$ days and $\tau=\sim60-80$ days, and therefore, they concluded that the outburst has similar rise time and fall time.

We refit the same data using the GE model.
We reduced the ZTF and ATLAS data and generated host-subtracted LCs following \citet{Wang2024_AT2018gn}.
We fit the data simultaneously instead of fitting ZTF and ATLAS data separately as \citet{Lu2025_Mrk1018} did.
We fit by the Monte-Carlo Markov-Chain (MCMC) method with the code ${\it emcee}$ assuming log-uniformly prior distributions for $\sigma$ and $\tau$.
The posterior probability density distributions of $\sigma$ and $\tau$ are displayed in Figure~\ref{fig:mrk1018_optlc}(a).
As can be seen from the figure, the observed data can be reproduced by two different groups of parameters, which can roughly be divided by the threshold of $\sigma=60$ day.
For one group $\sigma=33\pm14$ day and $\tau=116\pm6$ day, representing fast rising models with shorter rise time than fall time (Figure~\ref{fig:mrk1018_optlc}(b)).
And for the other group $\sigma=98^{+8}_{-11}$ day and $\tau=83^{+14}_{-10}$ day, representing slowly rising models with similar rise time and fall time (Figure~\ref{fig:mrk1018_optlc}(c)).
The integral posterior probabilities for the two groups are 17\% and 83\%, respectively, indicating that slowly rising models have a higher probability, but fast-rising models are also acceptable by the data.
Taking into account systematic errors due to AGN intrinsic variability, the differences between the two groups of models would be even smaller.
Therefore, we concluded that the LC data alone cannot rule out the possibility of UV LCs with fast rise and slow fall.

\begin{figure}
\centering
 \includegraphics[scale=0.85]{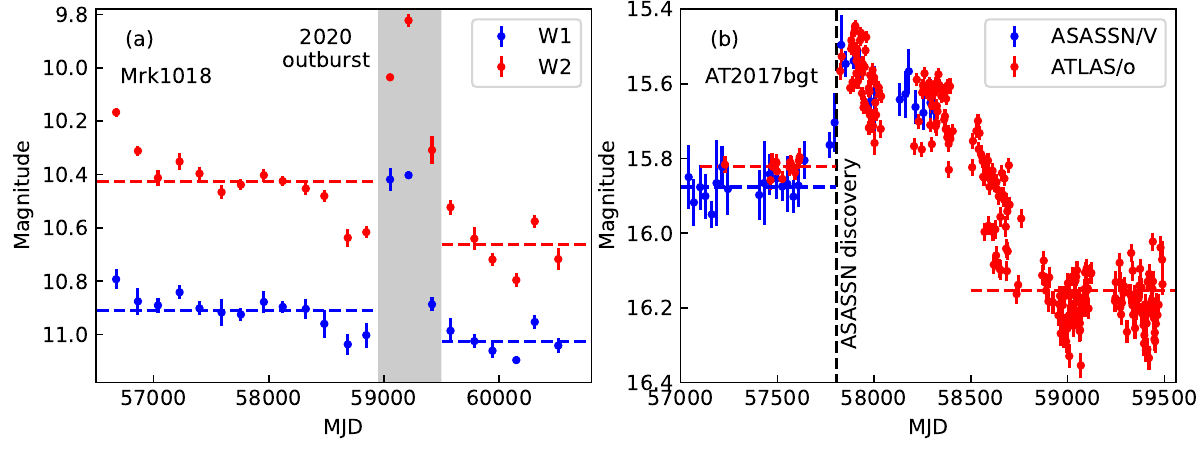}
 \caption{
  \textbf{(a)}: WISE MIR LC of Mrk 1018, and the levels before and after the 2020 outburst in dashed lines in W1 (blue) and W2 (red) bands.
  \textbf{(b)}: ASASSN/V and ATLAS/o LCs of AT 2017bgt.
  We labeled the discovery time of the flare and the levels before and after the flare.
  }
\label{fig:mrk1018_at2017bgt}
\end{figure}

A rapid outburst occurred during a long-term turning-off changing-look event in Mrk 1018.
As such rapid outbursts are rare, the outburst and the changing-look event are likely physically related \citep{Lu2025_Mrk1018}.
Here, we present a similar case AT 2017bgt, an outburst in a type 1 AGN \citep{Trakhtenbrot2019}.
As shown in Figure~\ref{fig:mrk1018_at2017bgt}, after the outburst, the optical flux dropped to a lower level than before the outburst, suggesting an underlying long-term dimming trend of the AGN, similar to Mrk 1018.
This similarity may imply a similar physical origin; hence, the outburst in Mrk 1018 may have similar properties to AT 2017bgt.
Thus, for our study of color variation of the MIR outbursts, it is uncertain whether Mrk 1018 should be classified as CLAGN or ANT.

\section{The parameters of MIR color variation for the MIR selected samples} \label{appendix:param_mir}

\setcounter{table}{0}
\renewcommand{\thetable}{D\arabic{table}}
\renewcommand*{\theHtable}{\thetable}

This section lists the CVR, IC and probabilities of falling into U, M and L regions.
The parameters of the MIRONGs hosted in non-Seyfert and Seyfert galaxies are listed in Table~\ref{tab:cvr_ic_nonseyfert} and Table~\ref{tab:cvr_ic_seyfert}, respectively, and those of the WTP sample are listed in Table~\ref{tab:cvr_ic_m24}.

\begin{longtable}{*8{c}}
\caption{CVR and IC of the MIRONGs in non-Seyfert galaxies.} \\
\hline
\hline
Object      & Redshift & subsample   & CVR               & IC               & $p_U$ & $p_M$ & $p_L$ \\
\hline
\endhead
\hline
\endfoot
\hline
SDSS J0045-0047 & 0.05677  & Star-forming & $0.04 \pm 0.22 $  & $0.93 \pm 0.31 $ & 4.85\%  & 34.35\%       & 60.80\%   \\
SDSS J0103+1401 & 0.04181  & Star-forming & $-0.07 \pm 0.03 $ & $1.04 \pm 0.02 $ & 0.00\%  & 0.00\%        & 100.00\%  \\
SDSS J0121+1405 & 0.12938  & Star-forming & $0.43 \pm 0.66 $  & $0.75 \pm 0.23 $ & 51.58\% & 17.27\%       & 31.16\%   \\
SDSS J0205+0004 & 0.07649  & Star-forming & $0.31 \pm 0.11 $  & $0.71 \pm 0.04 $ & 20.89\% & 76.80\%       & 2.31\%    \\
SDSS J0757+1908 & 0.10501  & Composite    & $0.33 \pm 0.23 $  & $0.48 \pm 0.32 $ & 38.26\% & 45.42\%       & 16.32\%   \\
SDSS J0814+2611 & 0.07567  & Star-forming & $0.54 \pm 0.41 $  & $0.78 \pm 0.32 $ & 62.98\% & 22.58\%       & 14.45\%   \\
SDSS J0837+4143 & 0.09806  & Composite    & $1.45 \pm 0.80 $  & $0.54 \pm 0.29 $ & 90.63\% & 4.81\%        & 4.56\%    \\
SDSS J0847+5142 & 0.11997  & Star-forming & $0.36 \pm 1.05 $  & $1.20 \pm 0.29 $ & 48.32\% & 11.52\%       & 40.17\%   \\
SDSS J0859+0922 & 0.15188  & Star-forming & $0.07 \pm 0.06 $  & $0.59 \pm 0.08 $ & 0.00\%  & 32.89\%       & 67.11\%   \\
SDSS J0909+1920 & 0.10716  & Composite    & $0.21 \pm 0.14 $  & $0.89 \pm 0.10 $ & 9.33\%  & 69.00\%       & 21.68\%   \\
SDSS J0915+4814 & 0.10049  & LINER        & $0.12 \pm 0.27 $  & $0.56 \pm 0.39 $ & 14.06\% & 38.04\%       & 47.90\%   \\
SDSS J0931+6626 & 0.08729  & LINER        & $-0.05 \pm 0.28 $ & $1.06 \pm 0.35 $ & 4.83\%  & 24.04\%       & 71.13\%   \\
SDSS J0943+5958 & 0.07491  & LINER        & $0.27 \pm 0.83 $  & $1.23 \pm 0.24 $ & 43.94\% & 14.16\%       & 41.90\%   \\
SDSS J0944+3105 & 0.03465  & Composite    & $0.28 \pm 0.22 $  & $0.79 \pm 0.19 $ & 29.17\% & 49.85\%       & 20.98\%   \\
SDSS J0957+0207 & 0.12528  & Composite    & $0.54 \pm 0.54 $  & $0.56 \pm 0.24 $ & 59.47\% & 19.19\%       & 21.34\%   \\
SDSS J1001+1829 & 0.10603  & Star-forming & $0.44 \pm 0.52 $  & $0.59 \pm 0.19 $ & 52.00\% & 21.75\%       & 26.25\%   \\
SDSS J1002+4424 & 0.15446  & LINER        & $-0.10 \pm 0.70 $ & $1.60 \pm 0.25 $ & 23.95\% & 15.02\%       & 61.03\%   \\
SDSS J1011+5348 & 0.2344   & LINER        & $0.22 \pm 0.14 $  & $0.24 \pm 0.25 $ & 8.64\%  & 71.01\%       & 20.35\%   \\
SDSS J1020+2515 & 0.13145  & Star-forming & $0.73 \pm 0.91 $  & $0.60 \pm 0.32 $ & 63.75\% & 11.48\%       & 24.77\%   \\
SDSS J1037+3912 & 0.10677  & Star-forming & $0.31 \pm 0.18 $  & $0.94 \pm 0.12 $ & 30.76\% & 56.00\%       & 13.24\%   \\
SDSS J1043+2716 & 0.12812  & Composite    & $0.04 \pm 0.09 $  & $0.73 \pm 0.16 $ & 0.00\%  & 25.14\%       & 74.86\%   \\
SDSS J1053+5524 & 0.15174  & Star-forming & $1.20 \pm 0.74 $  & $0.94 \pm 0.17 $ & 85.49\% & 7.34\%        & 7.17\%    \\
SDSS J1058+5444 & 0.13062  & Star-forming & $0.54 \pm 0.48 $  & $0.88 \pm 0.19 $ & 61.42\% & 20.53\%       & 18.05\%   \\
SDSS J1109+3708 & 0.02602  & LINER        & $0.08 \pm 0.12 $  & $0.81 \pm 0.20 $ & 0.41\%  & 44.02\%       & 55.57\%   \\
SDSS J1111+5923 & 0.16973  & Composite    & $0.51 \pm 0.45 $  & $0.50 \pm 0.13 $ & 59.48\% & 22.11\%       & 18.41\%   \\
SDSS J1115+0544 & 0.08995  & LINER        & $0.41 \pm 0.08 $  & $0.68 \pm 0.06 $ & 54.83\% & 45.16\%       & 0.01\%    \\
SDSS J1120+1933 & 0.1278   & Composite    & $0.62 \pm 0.45 $  & $0.34 \pm 0.21 $ & 69.23\% & 18.73\%       & 12.04\%   \\
SDSS J1124+0455 & 0.07398  & Composite    & $-0.04 \pm 0.20 $ & $1.41 \pm 0.15 $ & 1.70\%  & 23.59\%       & 74.71\%   \\
SDSS J1129+5131 & 0.03286  & Composite    & $0.38 \pm 0.70 $  & $1.53 \pm 0.23 $ & 48.88\% & 16.67\%       & 34.45\%   \\
SDSS J1139+6134 & 0.13461  & Star-forming & $0.07 \pm 0.15 $  & $0.96 \pm 0.27 $ & 1.37\%  & 41.38\%       & 57.25\%   \\
SDSS J1152+4850 & 0.15102  & Composite    & $-0.08 \pm 0.12 $ & $0.83 \pm 0.25 $ & 0.01\%  & 6.96\%        & 93.03\%   \\
SDSS J1153+4037 & 0.1451   & Star-forming & $-0.02 \pm 0.26 $ & $0.79 \pm 0.11 $ & 5.57\%  & 27.27\%       & 67.15\%   \\
SDSS J1203+5859 & 0.04692  & Star-forming & $0.26 \pm 0.44 $  & $1.02 \pm 0.15 $ & 37.46\% & 26.29\%       & 36.26\%   \\
SDSS J1218+2951 & 0.13559  & Composite    & $0.14 \pm 0.07 $  & $0.79 \pm 0.15 $ & 0.01\%  & 69.54\%       & 30.45\%   \\
SDSS J1219+0516 & 0.08251  & Composite    & $0.11 \pm 0.17 $  & $0.66 \pm 0.25 $ & 3.83\%  & 47.76\%       & 48.42\%   \\
SDSS J1242+2537 & 0.08789  & Composite    & $0.15 \pm 0.48 $  & $0.79 \pm 0.23 $ & 29.70\% & 23.66\%       & 46.64\%   \\
SDSS J1245-0147 & 0.21543  & LINER        & $0.48 \pm 0.44 $  & $0.61 \pm 0.15 $ & 57.46\% & 23.35\%       & 19.19\%   \\
SDSS J1305+3953 & 0.07249  & LINER        & $0.57 \pm 0.34 $  & $0.61 \pm 0.13 $ & 69.45\% & 22.48\%       & 8.07\%    \\
SDSS J1310+2518 & 0.16039  & Composite    & $0.34 \pm 0.31 $  & $0.37 \pm 0.25 $ & 42.75\% & 35.47\%       & 21.78\%   \\
SDSS J1322+3301 & 0.1269   & LINER        & $-0.21 \pm 0.64 $ & $0.92 \pm 0.23 $ & 16.50\% & 14.30\%       & 69.20\%   \\
SDSS J1329+2341 & 0.07171  & Composite    & $0.24 \pm 0.11 $  & $0.98 \pm 0.16 $ & 7.99\%  & 81.20\%       & 10.82\%   \\
SDSS J1341-0049 & 0.17538  & Composite    & $0.41 \pm 0.36 $  & $0.17 \pm 0.29 $ & 50.24\% & 29.49\%       & 20.27\%   \\
SDSS J1352+0009 & 0.16596  & LINER        & $0.42 \pm 0.40 $  & $0.32 \pm 0.25 $ & 52.59\% & 26.29\%       & 21.12\%   \\
SDSS J1409+1057 & 0.05972  & LINER        & $0.93 \pm 0.65 $  & $0.35 \pm 0.24 $ & 79.56\% & 10.46\%       & 9.98\%    \\
SDSS J1412+4114 & 0.1025   & Composite    & $-0.27 \pm 0.58 $ & $0.87 \pm 0.19 $ & 12.61\% & 13.75\%       & 73.64\%   \\
SDSS J1422+0609 & 0.05636  & Composite    & $0.28 \pm 0.10 $  & $0.28 \pm 0.23 $ & 12.17\% & 83.49\%       & 4.34\%    \\
SDSS J1424+6249 & 0.10913  & Composite    & $0.19 \pm 0.11 $  & $0.55 \pm 0.14 $ & 3.76\%  & 75.45\%       & 20.79\%   \\
SDSS J1440+1758 & 0.11574  & Composite    & $0.29 \pm 0.59 $  & $0.58 \pm 0.18 $ & 42.85\% & 19.70\%       & 37.45\%   \\
SDSS J1442+5558 & 0.07689  & Composite    & $0.00 \pm 0.12 $  & $0.62 \pm 0.26 $ & 0.04\%  & 19.38\%       & 80.58\%   \\
SDSS J1447+4023 & 0.13025  & LINER        & $0.49 \pm 0.14 $  & $0.37 \pm 0.05 $ & 73.95\% & 25.82\%       & 0.22\%    \\
SDSS J1448+1137 & 0.06657  & Star-forming & $-0.15 \pm 0.43 $ & $1.15 \pm 0.38 $ & 9.40\%  & 17.88\%       & 72.72\%   \\
SDSS J1508+2602 & 0.08255  & Composite    & $0.18 \pm 0.26 $  & $1.26 \pm 0.18 $ & 19.60\% & 42.15\%       & 38.26\%   \\
SDSS J1512+2809 & 0.11552  & LINER        & $0.23 \pm 0.25 $  & $0.86 \pm 0.15 $ & 25.70\% & 43.96\%       & 30.34\%   \\
SDSS J1513+3111 & 0.07181  & Composite    & $0.65 \pm 0.22 $  & $0.45 \pm 0.10 $ & 86.13\% & 13.05\%       & 0.82\%    \\
SDSS J1524+5314 & 0.08513  & Star-forming & $0.54 \pm 0.26 $  & $0.36 \pm 0.09 $ & 70.91\% & 24.79\%       & 4.30\%    \\
SDSS J1541+0718 & 0.16305  & Star-forming & $0.20 \pm 0.17 $  & $0.80 \pm 0.28 $ & 13.25\% & 59.63\%       & 27.12\%   \\
SDSS J1548+2208 & 0.03127  & Composite    & $-0.08 \pm 0.03 $ & $1.09 \pm 0.07 $ & 0.00\%  & 0.00\%        & 100.00\%  \\
SDSS J1554+3629 & 0.23683  & Composite    & $0.11 \pm 0.14 $  & $0.39 \pm 0.23 $ & 1.95\%  & 50.25\%       & 47.80\%   \\
SDSS J1555+2120 & 0.07094  & Composite    & $0.05 \pm 0.09 $  & $0.94 \pm 0.20 $ & 0.01\%  & 30.16\%       & 69.83\%   \\
SDSS J1600+4612 & 0.19742  & Star-forming & $0.51 \pm 0.46 $  & $0.69 \pm 0.13 $ & 59.50\% & 21.89\%       & 18.60\%   \\
SDSS J1612+1416 & 0.072    & Composite    & $0.08 \pm 0.19 $  & $1.02 \pm 0.33 $ & 4.48\%  & 40.30\%       & 55.23\%   \\
SDSS J1628+4810 & 0.12454  & Star-forming & $1.14 \pm 0.80 $  & $0.69 \pm 0.27 $ & 82.26\% & 8.09\%        & 9.65\%    \\
SDSS J1632+4416 & 0.05789  & LINER        & $0.24 \pm 0.27 $  & $0.76 \pm 0.11 $ & 26.41\% & 41.95\%       & 31.64\%   \\
SDSS J1647+3843 & 0.08547  & Star-forming & $0.27 \pm 0.07 $  & $0.64 \pm 0.05 $ & 3.43\%  & 95.58\%       & 0.98\%    \\
SDSS J2146+1041 & 0.16358  & Star-forming & $0.54 \pm 0.46 $  & $0.29 \pm 0.18 $ & 61.14\% & 20.93\%       & 17.93\%   \\
SDSS J2203+1124 & 0.18627  & Composite    & $0.18 \pm 0.13 $  & $0.45 \pm 0.25 $ & 4.11\%  & 70.85\%       & 25.04\%   \\
SDSS J2215-0107 & 0.04775  & Composite    & $0.25 \pm 0.19 $  & $0.60 \pm 0.35 $ & 21.84\% & 57.34\%       & 20.81\%   \\
SDSS J2312+1335 & 0.16553  & Composite    & $0.55 \pm 0.51 $  & $0.59 \pm 0.18 $ & 61.69\% & 19.49\%       & 18.83\%
\label{tab:cvr_ic_nonseyfert}
\end{longtable}

\begin{table}
\caption{CVR and IC of the MIRONGs in Seyfert galaxies.}
\begin{tabular}{ccccccc}
\hline
\hline
Object      & Redshift  & CVR               & IC               & $p_U$ & $p_M$ & $p_L$ \\
\hline
SDSS J0000+1438 & 0.1366   & $0.30 \pm 0.21 $  & $0.52 \pm 0.25 $ & 31.86\%  & 51.02\%       & 17.11\%\\
SDSS J0027+0713 & 0.13109  & $0.29 \pm 0.48 $  & $0.65 \pm 0.38 $ & 41.01\%  & 24.30\%       & 34.69\%\\
SDSS J0158-0052 & 0.08044  & $0.36 \pm 0.06 $  & $0.28 \pm 0.14 $ & 26.68\%  & 73.32\%       & 0.00\% \\
SDSS J0745+2655 & 0.11481  & $0.08 \pm 0.30 $  & $0.89 \pm 0.11 $ & 14.76\%  & 33.47\%       & 51.77\%\\
SDSS J0811+4054 & 0.06704  & $0.04 \pm 0.43 $  & $0.70 \pm 0.17 $ & 20.38\%  & 24.02\%       & 55.60\%\\
SDSS J0841+0526 & 0.15631  & $0.27 \pm 0.12 $  & $0.64 \pm 0.12 $ & 14.79\%  & 76.34\%       & 8.87\% \\
SDSS J0854+1113 & 0.16719  & $0.07 \pm 0.35 $  & $0.48 \pm 0.30 $ & 17.26\%  & 29.41\%       & 53.33\%\\
SDSS J1003+0202 & 0.1247   & $0.45 \pm 0.33 $  & $0.48 \pm 0.10 $ & 55.48\%  & 29.56\%       & 14.96\%\\
SDSS J1008+1549 & 0.11765  & $0.07 \pm 0.15 $  & $0.24 \pm 0.22 $ & 1.35\%   & 40.54\%       & 58.11\%\\
SDSS J1009+3436 & 0.20863  & $-0.18 \pm 0.58 $ & $0.91 \pm 0.25 $ & 16.75\%  & 15.59\%       & 67.66\%\\
SDSS J1017+1224 & 0.10762  & $0.28 \pm 0.37 $  & $0.72 \pm 0.10 $ & 38.71\%  & 30.70\%       & 30.59\%\\
SDSS J1041+3412 & 0.14028  & $-0.20 \pm 0.81 $ & $0.59 \pm 0.30 $ & 23.82\%  & 12.73\%       & 63.44\%\\
SDSS J1051+2101 & 0.06593  & $-0.42 \pm 0.61 $ & $1.07 \pm 0.19 $ & 9.05\%   & 10.80\%       & 80.14\%\\
SDSS J1105+5941 & 0.03369  & $0.11 \pm 0.05 $  & $0.62 \pm 0.11 $ & 0.00\%   & 54.86\%       & 45.14\%\\
SDSS J1114+4056 & 0.15247  & $0.28 \pm 0.19 $  & $0.31 \pm 0.34 $ & 26.14\%  & 56.90\%       & 16.96\%\\
SDSS J1133+6701 & 0.03968  & $0.13 \pm 0.06 $  & $0.47 \pm 0.09 $ & 0.00\%   & 66.58\%       & 33.42\%\\
SDSS J1200+0648 & 0.03599  & $0.04 \pm 0.16 $  & $1.24 \pm 0.36 $ & 1.40\%   & 33.82\%       & 64.79\%\\
SDSS J1201+3525 & 0.19031  & $0.16 \pm 0.21 $  & $0.50 \pm 0.38 $ & 13.08\%  & 47.86\%       & 39.06\%\\
SDSS J1208+3305 & 0.28028  & $-0.02 \pm 0.27 $ & $0.71 \pm 0.31 $ & 5.52\%   & 26.73\%       & 67.75\%\\
SDSS J1238+0815 & 0.11378  & $0.09 \pm 0.13 $  & $0.70 \pm 0.26 $ & 0.85\%   & 44.57\%       & 54.58\%\\
SDSS J1308+0429 & 0.04832  & $-0.93 \pm 0.53 $ & $1.37 \pm 0.25 $ & 0.60\%   & 1.98\%        & 97.42\%\\
SDSS J1332+2036 & 0.11249  & $1.00 \pm 0.12 $  & $0.43 \pm 0.05 $ & 100.00\% & 0.00\%        & 0.00\% \\
SDSS J1340+1842 & 0.09018  & $0.12 \pm 0.22 $  & $0.75 \pm 0.25 $ & 9.58\%   & 43.62\%       & 46.80\%\\
SDSS J1341+1516 & 0.12553  & $0.21 \pm 0.27 $  & $0.78 \pm 0.21 $ & 23.89\%  & 42.39\%       & 33.72\%\\
SDSS J1402+3922 & 0.06375  & $1.08 \pm 0.18 $  & $0.21 \pm 0.08 $ & 99.99\%  & 0.01\%        & 0.00\% \\
SDSS J1430+2303 & 0.08105  & $0.01 \pm 0.12 $  & $0.68 \pm 0.24 $ & 0.04\%   & 22.56\%       & 77.40\%\\
SDSS J1504+0107 & 0.12826  & $0.14 \pm 0.16 $  & $0.52 \pm 0.33 $ & 5.00\%   & 54.47\%       & 40.53\%\\
SDSS J1511+2214 & 0.12048  & $-0.01 \pm 0.27 $ & $0.78 \pm 0.20 $ & 6.51\%   & 27.49\%       & 66.00\%\\
SDSS J1533+2729 & 0.07193  & $-0.18 \pm 0.21 $ & $0.85 \pm 0.18 $ & 0.32\%   & 8.90\%        & 90.77\%\\
SDSS J1537+5814 & 0.09356  & $0.10 \pm 0.16 $  & $0.44 \pm 0.22 $ & 2.97\%   & 47.90\%       & 49.12\%\\
SDSS J1556+4513 & 0.1808   & $0.22 \pm 0.17 $  & $0.52 \pm 0.12 $ & 14.28\%  & 61.72\%       & 24.00\%\\
SDSS J1657+2345 & 0.05914  & $0.13 \pm 0.02 $  & $0.79 \pm 0.04 $ & 0.00\%   & 94.32\%       & 5.68\% \\
SDSS J1659+2049 & 0.04513  & $0.51 \pm 0.58 $  & $0.96 \pm 0.22 $ & 57.64\%  & 18.75\%       & 23.62\%\\
SDSS J2115-0011 & 0.23285  & $0.55 \pm 0.70 $  & $0.29 \pm 0.21 $ & 58.56\%  & 15.59\%       & 25.85\%\\
SDSS J2141-0857 & 0.08729  & $0.21 \pm 0.17 $  & $0.89 \pm 0.20 $ & 12.77\%  & 60.16\%       & 27.07\%\\
SDSS J2150-0106 & 0.08791  & $0.29 \pm 0.05 $  & $0.35 \pm 0.06 $ & 2.23\%   & 97.76\%       & 0.01\% \\
SDSS J2156+0041 & 0.05389  & $0.44 \pm 0.64 $  & $1.18 \pm 0.26 $ & 52.74\%  & 17.75\%       & 29.51\%\\
SDSS J2310+2220 & 0.07829  & $-0.37 \pm 0.81 $ & $1.03 \pm 0.29 $ & 17.11\%  & 11.04\%       & 71.85\%\\
\hline
\end{tabular}
\begin{tablenotes}
   \item
\end{tablenotes}
\label{tab:cvr_ic_seyfert}
\end{table}

\begin{table}
%\centering
\caption{CVR and IC of the WTP sample.}
\begin{tabular}{cccccccc}
\hline
\hline
Object      & Redshift & subsample   & CVR               & IC               & $p_U$ & $p_M$ & $p_L$ \\
\hline
WTP14abnpgk & 0.02035  & Silver & $0.14 \pm 0.22 $  & $0.81 \pm 0.18 $ & 11.87\%  & 44.56\%       & 43.57\%   \\
WTP14acnjbu & 0.026    & Silver & $0.12 \pm 0.17 $  & $1.20 \pm 0.15 $ & 5.64\%   & 49.81\%       & 44.55\%   \\
WTP14adbwvs & 0.02489  & Silver & $0.23 \pm 0.24 $  & $1.33 \pm 0.18 $ & 23.62\%  & 45.99\%       & 30.39\%   \\
WTP14adeqka & 0.01895  & Gold   & $0.11 \pm 0.10 $  & $1.04 \pm 0.23 $ & 0.14\%   & 52.66\%       & 47.20\%   \\
WTP15abymdq & 0.03742  & Gold   & $0.19 \pm 0.22 $  & $0.21 \pm 0.34 $ & 16.94\%  & 49.27\%       & 33.80\%   \\
WTP15acbgpn & 0.03369  & Silver & $0.11 \pm 0.05 $  & $0.61 \pm 0.11 $ & 0.00\%   & 56.31\%       & 43.69\%   \\
WTP16aatsnw & 0.04513  & Silver & $0.54 \pm 0.57 $  & $0.95 \pm 0.22 $ & 59.84\%  & 18.28\%       & 21.88\%   \\
WTP17aaldjb & 0.04832  & Gold   & $-0.92 \pm 0.52 $ & $1.36 \pm 0.25 $ & 0.59\%   & 2.02\%        & 97.39\%   \\
WTP17aalzpx & 0.03725  & Gold   & $0.17 \pm 0.18 $  & $0.77 \pm 0.26 $ & 9.61\%   & 56.06\%       & 34.34\%   \\
WTP17aamoxe & 0.042    & Gold   & $0.88 \pm 0.07 $  & $0.45 \pm 0.03 $ & 100.00\% & 0.00\%        & 0.00\%    \\
WTP17aamzew & 0.03465  & Gold   & $0.28 \pm 0.22 $  & $0.78 \pm 0.19 $ & 29.44\%  & 49.94\%       & 20.63\%   \\
WTP17aanbso & 0.04692  & Gold   & $0.30 \pm 0.45 $  & $1.01 \pm 0.15 $ & 41.46\%  & 25.82\%       & 32.73\%   \\
WTP18aajkmk & 0.0287   & Gold   & $0.08 \pm 0.05 $  & $0.57 \pm 0.04 $ & 0.00\%   & 39.22\%       & 60.78\%   \\
WTP18aamced & 0.03672  & Silver & $1.09 \pm 0.62 $  & $0.35 \pm 0.32 $ & 86.50\%  & 7.86\%        & 5.64\%    \\
WTP18aampwj & 0.0375   & Gold   & $0.32 \pm 0.04 $  & $0.51 \pm 0.05 $ & 1.56\%   & 98.45\%       & 0.00\%    \\
\hline
\end{tabular}
\begin{tablenotes}
   \item
\end{tablenotes}
\label{tab:cvr_ic_m24}
\end{table}

\section{Simulations of dust echoes} \label{appendix:simulation}

\setcounter{figure}{0}
\setcounter{equation}{0}
\renewcommand{\thefigure}{E\arabic{figure}}
\renewcommand*{\theHfigure}{\thefigure}

\subsection{Simulation of echo of torus dust} \label{appendix:simu_torus}

\begin{figure}
\centering
 \includegraphics[scale=0.55]{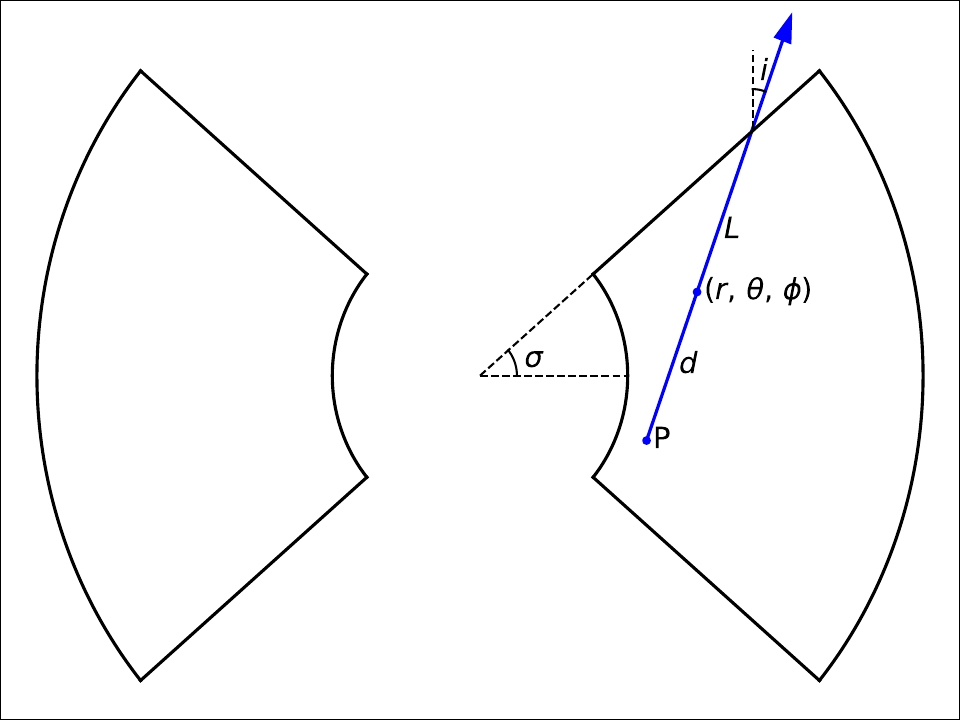}
 \caption{Illustration of the torus geometry in the simulation.}
\label{fig:torus}
\end{figure}

In this subsection, we introduce how we modified Lu16's model to calculate the echo of torus dust.

The dust geometry is shown in Figure~\ref{fig:torus}.
We required that the position $(R,\Theta,\Phi)$ of the dust satisfy $r_{\rm in}<R<r_{\rm out}$ and $\frac{\pi}{2}-\sigma<\Theta<\frac{\pi}{2}-\sigma$, where $r_{\rm in}$ and $r_{\rm out}$ are the inner and outer radii, and $\sigma$ is the half opening angle of the torus.
The torus is viewed with an inclination angle $i$, the angle between the line of sight and the torus axis.
In this work, we assumed $\sigma=40^\circ$, and hence the corresponding covering factor of the torus is 0.65.

The approaches to calculate the echo of torus dust are largely the same as those used to calculate the echo of spherical shell by Lu16.
The main differences lie in the following two points.
One is that when calculating the total luminosity via volume integration (equation 16 in Lu16), the lower and upper limits of $\theta$ are $\frac{\pi}{2}-\sigma$ and $\frac{\pi}{2}-\sigma$, respectively, instead of 0 and $\pi$.
The other is that when calculating the IR emergent optical depth for a point $P$ at a time $t$, instead of using equations 20 and 21 in Lu16, we used the following path integral:
\begin{equation}
\tau_{\rm IR}(P,t) = \int_L \pi a^2(r,t_r) n_d Q_{\rm IR} \Delta(r,\theta) {\rm d}l.
\end{equation}
In this equation, the path $L$ follows the line of sight.
For an infinitesimal path ${\rm d}l$ at a position of $(r, \theta, \phi)$, $a(r, t_r)$ is the grain size at the position at a delayed time $t_r=t+d/c$, where d is the distance to $P$, $\Delta(r,\theta)$ is a function reflecting whether the position is inside or outside the torus:
\begin{equation}
\Delta(r,\theta) = \left\{
\begin{array}{lr}
 1,\ \ \ \ {\rm if}\ r_{\rm in}<r<{\rm out}, \frac{\pi}{2}-\sigma<\theta<\frac{\pi}{2}-\sigma \\
 0,\ \ \ \ {\rm else}
\end{array}
\right.
\end{equation}

\subsection{Prediction on variation of NIR color} \label{appendix:simu_nir}

\begin{figure*}
\centering
 \includegraphics[scale=0.66]{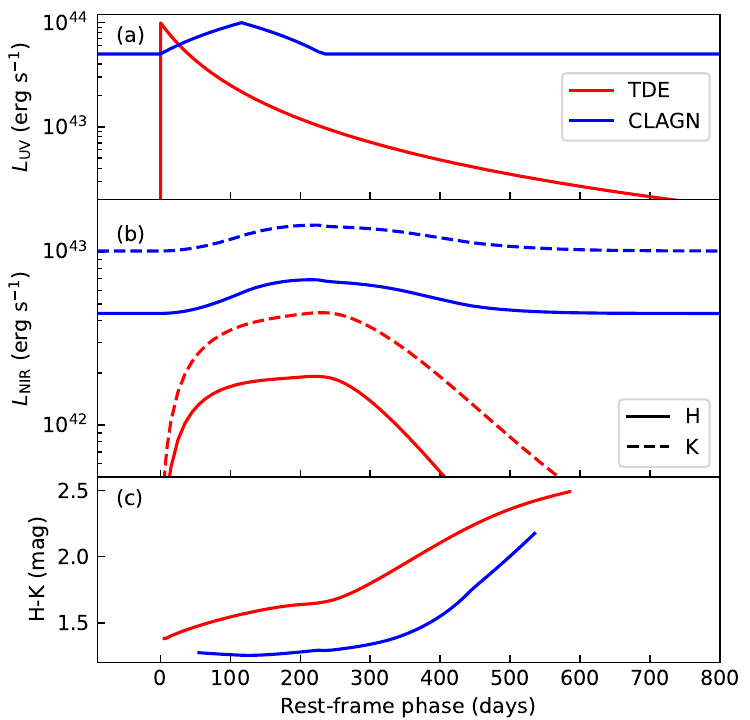}
 \caption{
  \textbf{(a)}: The assumed UV LCs for TDE and CLAGN cases.
  \textbf{(b)}: The simulated NIR LCs in the H (solid) and K (dashed) bands.
  \textbf{(c)}: The inferred H-K color curves.}
\label{fig:simu_nir}
\end{figure*}

We studied the possible difference in the NIR color of TDEs and CLAGNs using simulations.
For the TDE case, we set a flare with TDE form UV LC with $L_{\rm max}=10^{44}$ erg s$^{-1}$.
While for the CLAGN case, we set an outburst with LINEAR form UV LC with $L_{\rm max}=5\times10^{43}$ erg s$^{-1}$ and an underlying AGN with $L_{\rm AGN}=5\times10^{43}$ erg s$^{-1}$.
The settings of other parameters were the same as in subsection \ref{sec:simu_uvshape}.
We calculated the NIR LCs in H and K bands and the variation of $H-K$ color.
As shown in Figure~\ref{fig:simu_nir}, the NIR color of TDEs turns red rapidly in the rising phase, while that of CLAGNs keeps nearly steady in the rising phase.
Our simulations suggest that the NIR color variation has the potential to distinguish TDEs from CLAGNs.

\end{appendix}

\end{document}